\documentclass[aps,prl,twocolumn,superscriptaddress,showpacs]{revtex4-1}

\usepackage{amsmath}
\usepackage{graphicx}
\usepackage{calc}
\usepackage{bm}

\begin{document}

\title{Topological semimetals: surface transport and spin effects.}

\author{E.V. Deviatov}
\email[E-mail:~]{dev@issp.ac.ru}

\affiliation{Institute of Solid State Physics of the Russian Academy of Sciences, Chernogolovka, Moscow District, 2 Academician Ossipyan str., 142432 Russia}
\affiliation{V.L. Ginzburg Research Centre for High-Temperature Superconductivity and Quantum Materials, P.N. Lebedev Physical Institute of RAS, Moscow 119991, Russia}

\date{\today}

\begin{abstract}
For the solid state physics, recent interest to topological systems is mostly connected with topological semimetals, in particular,  to Weyl ones as the most representative semimetal  type.   Like other topological materials, e.g. topological and Chern insulators, topological semimetals acquire topologically protected surface states with linear dispersion. In contrast to helical surface states in topological insulators,  the surface states are chiral for Weyl semimetals, similarly to Chern insulators, which  allows to consider  Weyl semimetals as the three-dimensional analog of the quantum Hall effect regime. Weyl semimetals are also interesting for spin-dependent effects, due to the  spin-momentum locking in the topological surface states.  For topological semimetals, the main problem of transport investigations is to reveal the surface states contribution in the material with gapless bulk spectrum. Here, we present review of experimental results on charge and spin transport in topological semimetals: charge transport in   different superconducting  proximity devices; spin-dependent transport; magnetic response of the topological surface states; non-linear anomalous Hall effect as the direct manifestation of the non-zero Berry curvature  in topological semimetals. Possible applications are also considered for this new class of topological materials. 
\end{abstract}


\maketitle

\section{Introduction. Topological semimetals} 

As it is often the case in physics, terms that were once introduced can change their meaning, sometimes quite significantly. For example, in modern solid-state physics, the linear spectrum of charge carriers is commonly referred to as the Dirac spectrum. In contrast, the original definition of the Dirac spectrum in relativistic quantum mechanics was for a particle with a non-zero mass, while the linear spectrum was referred to as the Weyl one. Moreover, in solid-state physics, the term "Weyl spectrum" is now used to describe quite complicated systems known as Weyl semimetals. Even the meaning of the term "semimetal" has been changed. In classical textbooks~\cite{kittel}, semimetals are materials (such as bismuth, antimony, and others) where the valence band and conduction band overlap in energy, but they are separated in momentum space, so the bands do not intersect. Nowadays, topological semimetals are materials where there is not only spectrum  inversion, but also band crossing, with the degeneracy of energy levels determined by additional symmetry constraints~\cite{weyl_review}. In this sense, topological semimetals are closer to topological insulators than to the conventional semimetals from the classical textbooks~\cite{kittel}.

\subsection{Topological insulators}

\begin{figure}[t]
\center{\includegraphics[width=\columnwidth]{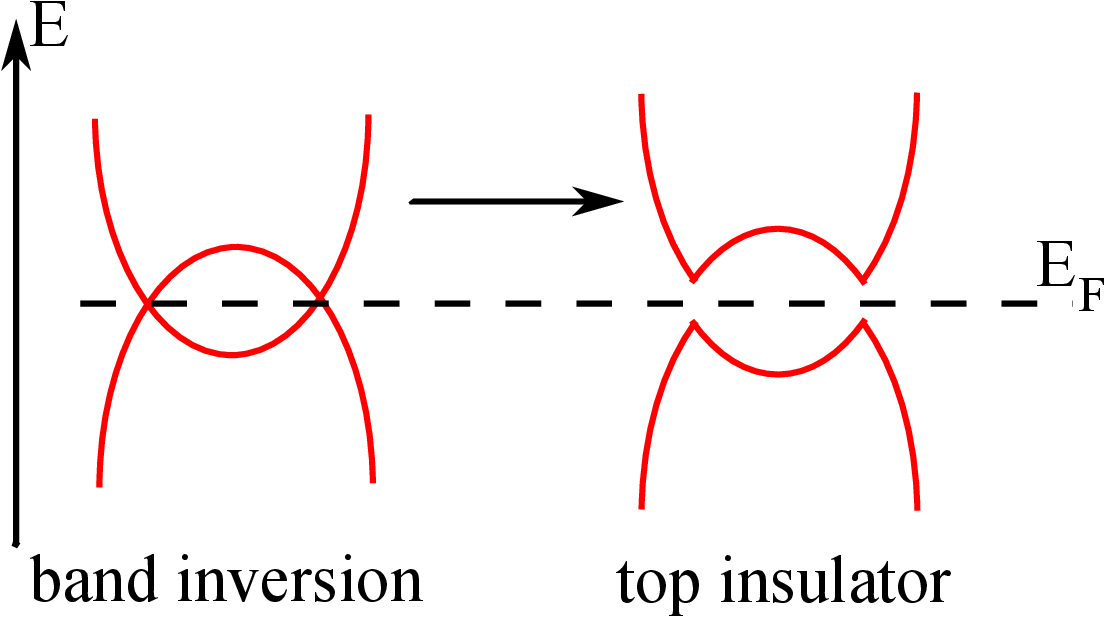}}
\caption{Schematic representation of the inversion of the bulk spectrum that occurs in some materials, such as Bi$_2$Te$_3$. In materials with band inversion, the hole band lies higher (in energy) than the electron one, precisely the opposite of the usual picture of the direct energy spectrum in a solid. Generally speaking, there should be a line of intersection of the bands in momentum space (a nodal line), so the spectrum of the inverse system could be expected to be gapless (left diagram in the figure). However, the states on the nodal line are doubly degenerate in energy, and as it is usual in perturbation theory, any weak perturbation lifts the degeneracy and leads to repulsion of the bands, that is, to the opening of an energy gap along the nodal line, as shown on the right. If the Fermi level is located in the opened gap, such a material will be an insulator from the standpoint of its bulk properties, and a topological insulator from the standpoint of edge (surface) effects.~\cite{PLAN_R40}}
\label{volkov}
\end{figure}

The main features of a topological insulator were predicted by Volkov and Pankratov in paper~\cite{PLAN_R40}, which was performed for the three-dimensional case long before widespread interest to this phenomenon. First of all, we are talking about systems with an inversion of the bulk spectrum, which arises in some materials (for example, Bi$_2$Te$_3$)~\cite{deviatov_top_ins}. The inversion itself can arise already in the tight-binding approximation~\cite{kittel} during the formation of a band structure from the levels of individual atoms: in this approximation, there is no a priori requirement to coincide for the original systematics of atomic levels and the order of the bands obtained from them. Everything is determined by the magnitude and sign of the corresponding constants~\cite{kittel}. In materials with band inversion, the hole band lies higher (in energy) than the electron one, exactly the opposite of the usual picture of the direct energy spectrum in a solid, see Fig.~\ref{volkov}. In this case, generally speaking, there should be a band intersection line in momentum space (a nodal line), so the spectrum of the inverse system might be expected to be gapless. However, the states on the nodal line are doubly degenerate in energy, and as it is typical in perturbation theory, any weak perturbation lifts the degeneracy and leads to band repulsion, i.e., to the opening of an energy gap along the nodal line (see Fig. ~\ref{volkov}). If the Fermi level is located in the gap, such a material will be an insulator from the standpoint of its bulk properties~\cite{kittel}.

The situation is different for edge effects. The authors of the work~\cite{PLAN_R40} considered the edge of a sample made from the material with band inversion as a smooth transition to a material with a direct-band (conventional, non-inverted) spectrum. Along such a transition, the bands obviously swap places, i.e., first the inverse gap decreases to zero, then the direct gap in the spectrum begins to increase. The law by which the gap changes depends on the specific materials and the structure of the boundary, but the very presence of a point of change in the type of spectrum near the boundary between the materials is determined only by the need to change the spectrum from inverted to direct ones, i.e., by  general, topological considerations. Near the band reconstruction point, an interface gapless state arises~\cite{PLAN_R40} (in the authors' terminology - Weyl state, in modern terminology - Dirac one), which is characterized by two spin-split oppositely directed modes. A special case of the problem considered in~\cite{PLAN_R40} is the contact of an inverse material with a vacuum: the physical boundary of the sample, the bulk spectrum of which is characterized by band inversion, is a potential barrier for both electrons and holes, which requires the spectrum to be reconstructed from the inverse to the direct one.

Thus, a topological insulator is characterized by an energy gap in the bulk spectrum and by topologically protected (at least in the sense of their existence) gapless edge states~\cite{kane2005,PLAN_R36}.

\subsection{Quantum Hall effect. Symmetry of the surface (edge) states}

\begin{figure}[t]
\center{\includegraphics[width=\columnwidth]{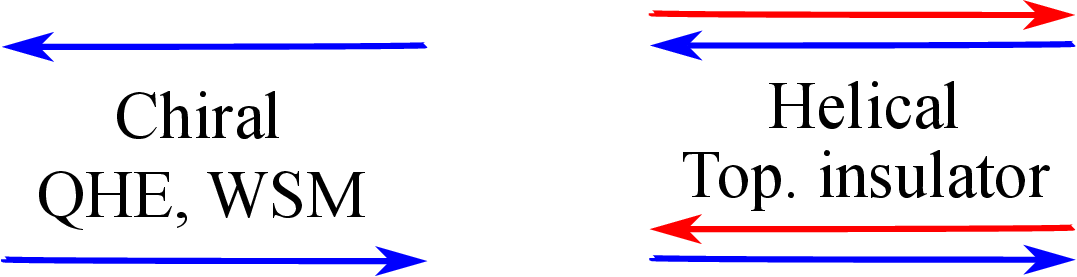}}
\caption{
Edge state symmetry. Chiral edge states of the integer quantum Hall effect (QHE)~\cite{review} (left) and helical edge states of a two-dimensional topological insulator~\cite{Molenkamp} (right). Each edge state of a topological insulator is a pair of QHE states propagating along the same edge of the sample in opposite directions~\cite{haldane} and, at the same time, spin-split due to the spin-momentum locking, as indicated by the different colors of the arrows. From the point of view of the edge state symmetry, a topological Weyl semimetal (WSM) is a generalization of the Chern insulator (the quantum Hall effect regime) to the three-dimensional case and zero external magnetic field~\cite{weyl_review}.}
\label{sample2p}
\end{figure}

Another example of a topologically nontrivial system is the quantum Hall effect regime, where chiral (i.e. with definite direction) edge states arise in a quantizing magnetic field, again in the absence of bulk conductivity, see review~\cite{review}. In this case, the gap in the bulk spectrum appears as a gap between the equidistant Landau levels, which, due to the presence of the edge potential, bend upward at the edges of the sample~\cite{buttiker88}. For an integer number of completely filled Landau levels, the Fermi level lies in the gap of the bulk spectrum, while at the edges, every of the filled Landau levels intersects the Fermi level. Thus, even with an energy gap in the bulk, there are delocalized gapless states at the edges. Such one-dimensional states, located  along the line of intersections of the Fermi level and Landau levels, were called current-carrying edge states~\cite{halperin}. Edge state transport allows one to describe both qualitatively and quantitatively the transport in the regime of the integer quantum Hall effect~\cite{buttiker88}. The direction of their propagation (the group velocity of electrons) is determined by the bending of the Landau level at the edge~\cite{buttiker88}. In other words, the propagation direction is defined by the normal to the edge and the direction of the magnetic field, which determines the chiral nature of edge states. The oppositely directed states are concentrated at different edges of the sample, see Fig.~\ref{sample2p}, it is this fact that determines the perfect protection from backscattering in the quantum Hall effect regime in macroscopic samples and the ideal quantization of the Hall resistance~\cite{buttiker88}.

In contrast to the quantum Hall effect regime, the edge states of a topological insulator arise in the absence of a magnetic field and have time-reversal symmetry~\cite{kane2005,PLAN_R36}. In the simplified picture, each edge state of a topological insulator is a pair of quantum Hall  states propagating along the same edge of the sample in opposite directions~\cite{haldane}, see Fig.~\ref{sample2p}, i.e. it is a non-directional helical, but spin-split edge state~\cite{Molenkamp}. In this case, the protection against backscattering ceases to be perfect: although the direction of motion of a charge carrier along the edge is rigidly related to the spin polarization of this carrier, i.e. backscattering must necessarily be accompanied by a change in spin, there is no prohibition on two-particle processes of simultaneous scattering of counter-propagating particles, which limits~\cite{glazman_mirlin,mirlin,glazman} the accuracy of resistance quantization~\cite{Molenkamp} in topological insulators, see more in the review~\cite{deviatov_top_ins}.

From the above, it is clear that the quantum Hall effect regime itself is possible only for two-dimensional systems, when the dimensional quantization in the potential well prohibits the movement of the charge carrier along the magnetic field, and there is a gap between the Landau levels in the bulk spectrum~\cite{review}. Topological insulators are possible in both two-dimensional and three-dimensional systems; the presence of a bulk gap here does not depend on the sample dimension and the most important thing is the presence of band inversion~\cite{kane2005,PLAN_R36}. 

From the point of view of the symmetry of topological surface states, a topological Weyl semimetal is a generalization of a Chern insulator (quantum Hall effect regime) to the three-dimensional case and zero external magnetic field~\cite{chern_3D_review}. This primarily raises the question of the origin of chiral surface states without an external magnetic field.

\subsection{Berry curvature. Topological invariants} \label{berry_section}

For further discussion, it will be useful to recall several features of the Bloch functions of electrons in a crystal~\cite{weyl_review}, which are not considered in classical textbooks~\cite{kittel} on solid state physics.

First of all, for a crystalline solid, the Bloch functions are invariant with respect to a local phase shift in momentum space. Specifically, it is easy to show by direct calculation that if $\psi_{n,k}(r)=u_{n,k}exp(ikr)$ is a Bloch function for an electron in a crystal, i.e., an eigenfunction of the electron's Hamiltonian in a periodic potential, then $\psi_{n,k}(r)=e^{i\phi(k)}u_{n,k}exp(ikr)$ is also an eigenfunction (Bloch function) for the same Hamiltonian, for any momentum-dependent phase $\phi(k)$.

In this case, one can define the quantity $A_n=-i<u_{n,k}|\nabla_k |u_{n,k}>$, which has the meaning of an effective vector potential in the momentum space: with a local phase shift $\phi_n(k)$ in the momentum space, $|u_{n,k}>$ transforms into $e^{i\phi_n(k)}|u_{n,k}>$, and $A_n$ transforms into $A_n+\nabla_k \phi_n(k)$. The effective magnetic field (Berry curvature) is introduced in the usual way as $F_n=\nabla_k \times A_n$ , it is its non-zero value that leads to the appearance of the Hall term $\sim eE\times F$ in the equation of electron motion in real space, and, ultimately, determines the topological non-triviality (the presence of chiral surface states) of a Weyl semimetal.

Further considerations can be carried out by analogy with the quantum Hall effect regime~\cite{tauless_iqhe}. The integral $$\oint_{BZ} Fdk=2\pi C$$ calculated over the Brullien zone for a nonzero and integer Chern number $C$ determines the quantization of the Hall conductivity. The Chern number $C$ in this sense was introduced by Thouless for two-dimensional electron systems in a magnetic field and is also known as the TKNN invariant~\cite{tauless_iqhe}. In the quantum Hall effect regime, it has a clear physical meaning: $C$ is equal to the number of intersection points of the Fermi level by the Landau levels at the edge of the sample, that is, the number of edge states. An integer value of $C$ corresponds to an integer number of filled Landau levels, i.e., dissipationless transport channels in the quantum Hall effect regime~\cite{buttiker88}. The normal insulator in this description corresponds to zero Chern number $C=0$.

For generality, it should be noted that for helical edge states in two-dimensional  topological insulators, only the parity of the number of Fermi-level crossing points at the edge of the sample matters, which is reflected in the topological invariant $Z_2$, which is defined analogous to the Chern number~\cite{kane2005}. The concept of the $Z_2$ invariant was subsequently extended (with inevitable mathematical complications) to three-dimensional systems~\cite{PLAN_R36}.

The main problem with this type of topological classification, despite its theoretical beauty, is its somewhat retrospective nature. In real systems, there are always electron-electron interactions, which are quite strong in general case. It is believed that the described classification can be extended to the case of electron-electron interaction in the quantum Hall effect regime, but this is currently not obvious for topological insulators and topological semimetals. Accordingly, the search for topological materials is conducted experimentally among systems with strong spin-orbit coupling and bulk spectrum inversion, primarily using ARPES (angle-resolved photoemission spectroscopy), see, for example, the papers~\cite{PLAN_R38,PLAN_R39} for topological insulators.

\subsection{Weyl semimetals} 

Depending on the type of bulk spectrum, nowadays we distinguish between Dirac semimetals, Weyl semimetals, chiral semimetals and semimetals with a nodal line~\cite{weyl_review}. The clearest, most refined understanding of the concept of a topological semimetal and the introduction of chiral surface states (Fermi arcs) is provided by a consideration of a Weyl semimetal.

\begin{figure}[t]
\center{\includegraphics[width=\columnwidth]{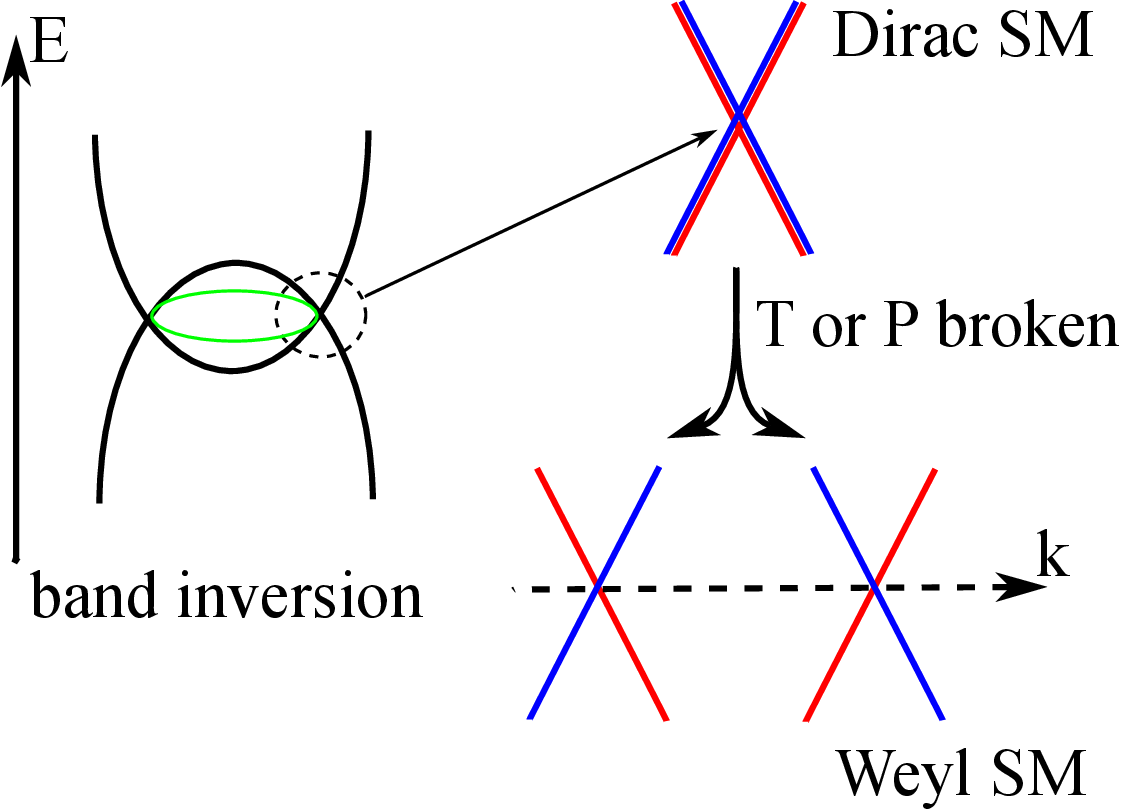}}
\caption{Formation of the bulk spectrum of a Weyl semimetal. Unlike a topological insulator (see Fig. ~\ref{volkov}), in the case of a topological semimetal, symmetry constraints imposed by the crystal lattice require the preservation of degeneracy at certain points on the nodal line (marked in red), which are then called Dirac points. In a small neighborhood of the Dirac points, the spectrum is linear and spin-degenerate, i.e., a Dirac semimetal (for example, Cd$_3$As$_2$, NiTi$_2$) is a three-dimensional analogue of graphene. For the Weyl spectrum, due to the violation of inversion symmetry P (e.g., in WTe$_2$) or time reversal symmetry T (e.g., in Co$_3$Sn$_2$S$_2$) in the crystal, each Dirac point, initially spin-degenerate, splits in the momentum space  into two Weyl nodes, which are at the same energy and characterized by different chirality (the color denotes the spin projection, the slope gives the sign of the velocity)~\cite{weyl_review} }
\label{bulk}
\end{figure}

As for a topological insulator, we start from a material with a band inversion, where a weak perturbation lifts the degeneracy and leads to the opening of an energy gap along the nodal line, as depicted in Fig.~\ref{volkov}. However, in the case of a topological semimetal, the system is subject to additional symmetry constraints imposed by the crystal lattice. In particular, symmetry requires preservation of degeneracy at some (in the simplest case of a rotation axis, two) points on the nodal line, called Dirac points, see Fig.~\ref{bulk}. The bulk spectrum of such a material (a Dirac semimetal in modern terminology) turns out to be gapless.  In essence, a Dirac semimetal is a three-dimensional analogue of graphene~\cite{weyl_review}.

We now require additionally a violation of time reversal symmetry or the inversion symmetry~\cite{weyl_review}. In this case, each Dirac point (initially spin-degenerate) splits in the momentum space  into two Weyl nodes, which are at the same energy and characterized by different chirality, see Fig.~\ref{bulk}. This bulk spectrum is characteristic of Weyl semimetals and leads to the appearance of chiral surface states (Fermi arcs).

\begin{figure}[t]
\center{\includegraphics[width=\columnwidth]{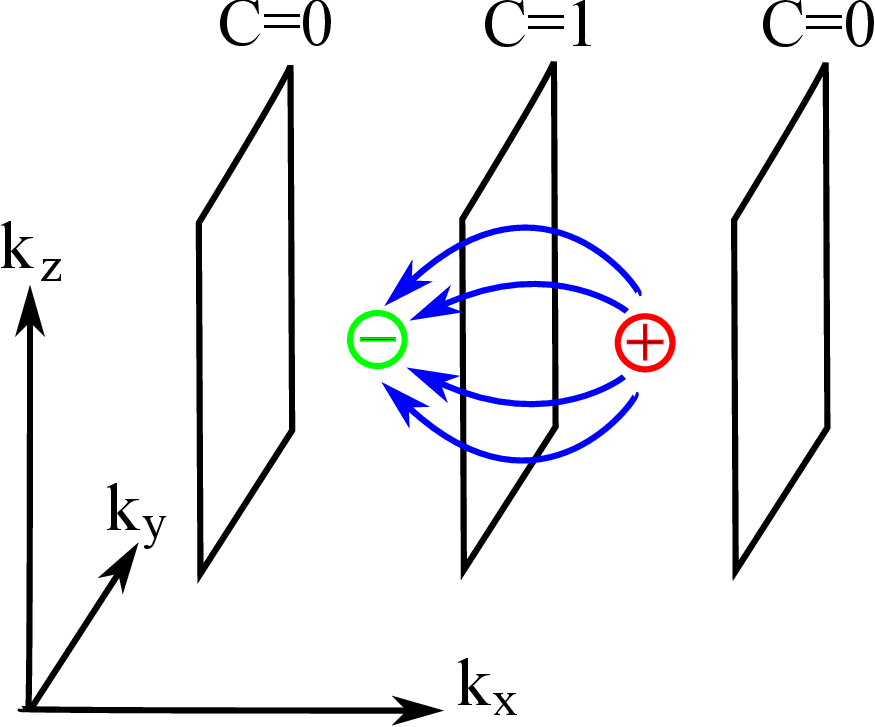}}
\caption{Representation of a three-dimensional Weyl semimetal as a stack of two-dimensional planes in momentum space~\cite{weyl_review}. For planes corresponding to $k_x$ values between the Weyl nodes, the Berry curvature is non-zero (the effective magnetic field $F_n=\nabla_k \times A_n$ is directed from one Weyl node to another). They are characterized by a non-zero Chern number $C \ne 0$. For each of these topologically non-trivial planes, one ($C=1$) or more (for $C>1$) chiral one-dimensional edge states should be introduced, as is done for the usual regime of the integer quantum Hall effect in two-dimensional systems. The sum of all one-dimensional chiral states forms a two-dimensional chiral surface state of the Fermi arc type on the crystal surface.}
\label{Fermi-arcs}
\end{figure}

There are several ways to introduce the surface states for Weyl semimetals~\cite{weyl_review}, let's consider one of them. The main problem when introducing surface states is the need to speculate in three-dimensional momentum space. In this case, it is convenient to reduce the problem to two dimensions by selecting one of the axes, for example, $k_x$. Based on the symmetry of the problem, the selected direction $k_x$ is parallel to the line between two adjacent Weyl nodes. In this case, one can represent the three-dimensional system as a set of two-dimensional planes ($k_y,k_z$) perpendicular to the selected axis, each of them is characterized by a certain value of the momentum $k_x$, see Fig.~\ref{Fermi-arcs}.

From the point of view of the Berry curvature introduced above, such planes can be divided into two types: (i) all planes corresponding to the values of $k_x$ between adjacent Weyl nodes with different chirality; for these planes the Berry curvature is non-zero. The effective magnetic field $F_n=\nabla_k \times A_n$ is directed from one Weyl node to another, the equation of electron motion in real space contains the Hall term $\sim eE\times F$, and, accordingly, all these planes are characterized by a non-zero Chern number $C \ne 0$; (ii) All other planes that lie outside this region of $k_x$ values; they are topologically trivial ($C=0$) due to the zero Berry curvature.

For each topologically nontrivial plane with a nonzero Chern number, one ($C=1$) or more (for $C>1$) chiral one-dimensional edge states should be introduced, as is done for the usual regime of the integer quantum Hall effect, with the Berry curvature acting as the effective magnetic field. Since we consider the original three-dimensional crystal as a set of two-dimensional planes in k-space, the sum of all one-dimensional chiral states of each topologically nontrivial plane forms a two-dimensional chiral surface state. In momentum space, it will be represented by a Fermi arc  connecting the projections of the Weyl nodes onto a given crystal surface~\cite{weyl_review}. On opposite surfaces, the Fermi arcs are directed toward each other, so that the dissipative-free (carried by the ground state) current encircles the sample.

\subsection{Concluding remarks on Weyl semimetals}

Several comments need to be made regarding the picture described above.

1. In the logic of this consideration, a Weyl semimetal is necessarily three-dimensional~\cite{weyl_review}: the Weyl nodes are separated by a finite distance in the direction perpendicular to the two-dimensional planes in Fig.~\ref{Fermi-arcs}. This statement well corresponds to experiment - it is well known that three-dimensional WTe$_2$ is one of the most prominent representatives of non-magnetic Weyl semimetals, but the monolayer WTe$_2$ is also a well-known two-dimensional topological insulator.

2. The Weyl semimetal introduced in this way is, from the point of view of its bulk properties, either magnetically ordered (a ferromagnet or an antiferromagnet if the Dirac point is split into two Weyl nodes due to the violation of time reversal symmetry), or a ferroelectric  if the symmetry with respect to the inversion center is broken~\cite{weyl_review,TSreview}. In any case, the material is characterized by a gapless bulk spectrum~\cite{weyl_review}, that leads to a noticeable bulk conductivity, which makes non-magnetic Weyl semimetals also representatives of the class of polar metals~\cite{PM,pm1,pm2,pm4}.

\begin{figure}[t]
\center{\includegraphics[width=\columnwidth]{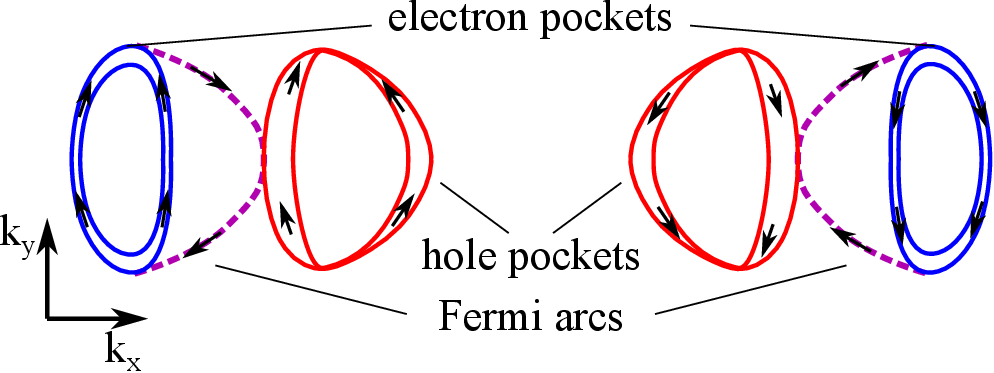}}
\caption{ Schematic diagram of Fermi arcs in the surface Brilliant zone and spin-polarized spin pockets in the bulk spectrum for the nonmagnetic Weyl semimetal WTe$_2$. In this material, due to the tilt of the Weyl cones (type II semimetal), bulk electrons and holes are present, similar to classical semimetals, along with counter-directed Fermi arcs on the surface connecting the projections of the Weyl nodes. The splitting line of the Weyl nodes coincides with the b-axis of the crystal, so the preferred direction of carrier propagation in counter-directed chiral arcs coincides with the a-direction in highly anisotropic WTe$_2$. Arrows indicate spin projections, which are defined by the Weyl surface states dispersion due to spin-momentum locking. Figure from the paper~\cite{WTe2_Ni} based on ARPES measurements~\cite{das16,feng2016}.}
\label{Fermi-arcs-ARPES}
\end{figure}

3. The above-described picture with a single chiral surface state connecting the projections of two Weyl nodes onto the surface and directed in opposite directions on opposite crystal faces implies a violation of time reversal symmetry (a circular diamagnetic current flows). In this simplest form, it can only be valid in the simplest and most idealized case -- a magnetic Weyl semimetal with one pair of Weyl nodes. In non-magnetic materials (with inversion symmetry breaking), counter-directed states arise, originating from (at least) two pairs of Weyl nodes, see Fig.~\ref{Fermi-arcs-ARPES} for WTe$_2$. In some real materials, the number of nodes can reach 32 (the TaAs family)~\cite{weyl_review}.

From the point of view of studying the physical properties of topological surface states, it is natural to use materials with a minimum number of Weyl node pairs. The detailed information about the structure of surface states can be obtained from ARPES~\cite{Xu} measurements, which essentially allow visualization of Fermi arcs and are the best confirmation of the validity of the picture presented above, for a lot of examples, see~\cite{weyl_review}. The figure Fig.~\ref{Fermi-arcs-ARPES} shows an example of a diagram of Fermi arcs on the surface Brilliant zone for the non-magnetic Weyl semimetal WTe$_2$, constructed~\cite{WTe2_Ni} from ARPES measurements~\cite{das16,feng2016}.

\section{Statement of the problem}

The main question for  experiment is whether it is possible to detect the contribution of Fermi arc topological states in transport, similar as the transport measurements allow us to study edge transport in the integer quantum Hall effect regime.

The transport experiment is complicated by the obligatory presence of bulk charge transfer due to the gapless bulk spectrum of Weyl semimetals, as well as by the need to separate the contribution of Fermi arc topological transport from trivial surface effects associated, for example, with the band bending at the surface of the sample~\cite{zav1,zav2,hofmann}.

In the literature on Weyl semimetals, two transport effects have already become classical: the anomalous Hall effect and the chiral anomaly. 

(i) For a Weyl semimetal, as for any conducting system, the usual Hall effect should be observed in an external magnetic field, i.e., the Hall resistance should be directly proportional to magnetic field. However, for a magnetic Weyl semimetal, the Hall signal does not disappear when the field is reduced to zero. Specifically, in a zero magnetic field, the Hall signal is determined by the dissipative (diamagnetic) current of the chiral surface states. This signal will change the sign when the bulk of the sample is re-magnetized by an external magnetic field. Thus,  a hysteresis loop is observed in the Hall resistance, which is called the anomalous Hall effect~\cite{weyl_review}. Formally speaking, the anomalous Hall effect is the most obvious manifestation of chiral surface transport in magnetic topological semimetals.

(ii) The chiral anomaly manifests itself as negative magnetoresistance of a topological semimetal in low magnetic fields. When the directions of the magnetic and electric fields (magnetic field and current through the sample) coincide, an imbalance of carriers at the zero Landau level in the Weyl nodes occurs, which leads to an increase in conductivity and, accordingly, to a drop in resistance in a magnetic field~\cite{weyl_review}. The effect obviously depends strongly on the angle between the field and the current, which underlies its experimental demonstration.

Both effects are macroscopic and therefore easily measurable, but do not provide convincing confirmation of the topological nature of the materials being studied. Solid state physics knows various mechanisms leading to negative magnetoresistance (for example, weak localization), as an example we can cite the recent work~\cite{chiral_wang}. As for the anomalous Hall effect, it can also arise in topologically trivial systems due to various types of spin-dependent scattering.

For these reasons, experiments are needed in which the contribution of transport from chiral surface states can be unambiguously identified.

\section{Experimental geometry}

For any transport experiments with topological semimetals, it is  necessary to define the desired contact configuration for a particular experiment. In this case, we cannot directly use methods developed for two-dimensional semiconductor systems: in the latter case, fabrication of the transport sample begins with mesa etching, by removing all layers down to the quantum well (typically around 200 nm). A Weyl semimetal is by definition a three-dimensional object, and attempting to define its shape by etching would result in approximately a micron thick  mesa step. Such a height difference is unsuitable for  any type of metal deposition (or even resists in lithography). One could attempt to use thick epitaxial films, while maintaining the three-dimensional regime, but this approach is more suitable for applications; for research purposes, it is highly desirable to work with  single crystals.

\begin{figure}[t]
\center{\includegraphics[width=\columnwidth]{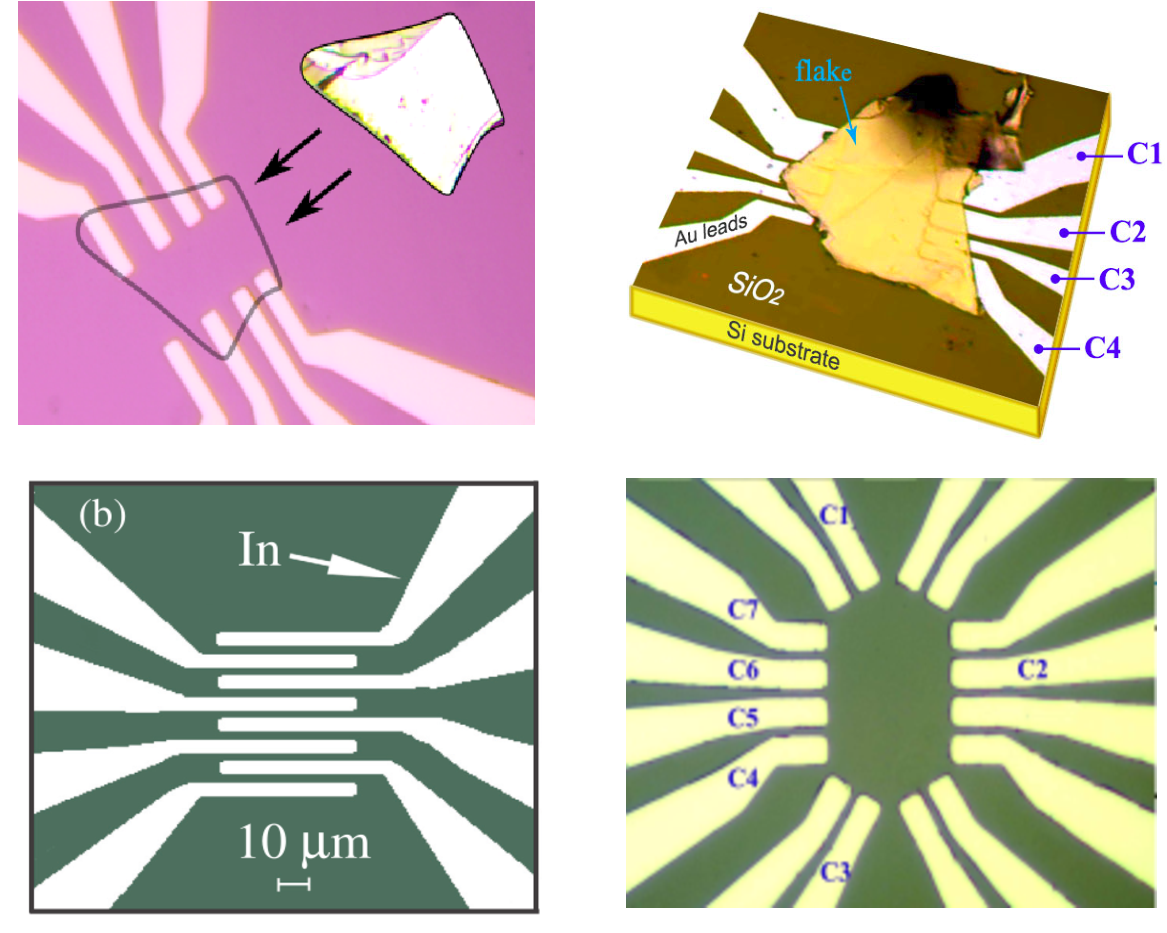}}
\caption{A method for fabricating a specified configuration of contacts to three-dimensional conducting flakes of topological semimetals. The desired contact configuration is created on the surface of oxidized silicon using standard lithographic techniques. Approximately 1~$\mu$m thick flake  of the material  is placed on the contacts and pressed once, as shown in the upper left. The final sample is shown in the upper right; the flake rests freely on the contacts. For most materials, the contacts can withstand repeated thermal cycling without degradation. Furthermore, the  surface of the crystal, pressed against the contact pattern, is protected from any environmental contamination by SiO2 substrate. Bottom row: various contact configurations, e.g.  for studying the Josephson effect (left) and the nonlinear anomalous Hall effect (right).}
\label{samples}
\end{figure}

However, to take full advantage of modern semiconductor technology, the process can be reversed, similar to what was done on the early stages of graphene research. Specifically, the desired contact configuration is created on the surface of oxidized silicon, i.e., a standard industrial silicon washer, by using standard methods - lithography, thin-film deposition (50-200 nm depending on the film material), and lift-off lithography (see Fig. ~\ref{samples}).

After fabricating the contact pattern, a thin flake of the material  is placed atop the contacts, as shown in Fig. ~\ref{samples}. The flake itself is obtained by mechanical exfoliation from a bulk single crystal, approximately 1~$\mu$m thick and approximately 100~$\mu$m in lateral dimensions. Most topological semimetals have a layered structure, and the exfoliation process is straightforward. The most perfect flakes (plane-parallel, with a minimum of visible surface defects) are selected after exfoliation. After placing the flake in the desired location on the contact pattern, it is pressed  once with a clean, oxidized silicon surface. A special metal frame is used to prevent accidental shifts and control the pressing force. After the process is complete, the pressing surface is removed, and the flake then rests freely on the contacts.

This technique enables the creation of high-quality normal (gold), superconducting (indium, niobium), and ferromagnetic (nickel, permalloy) contacts. In particular, the transparency of the resulting contacts, although varying from sample to sample, allows to investigate Josephson current on the surface of the Weyl semimetal in the case of superconducting contacts. For most materials, the contacts withstand repeated thermal cycling without degradation. As an additional advantage, the surface of the crystal with metallic contacts is protected from any environmental contamination by SiO2 substrate, which was tested even on such a sensitive material as black phosphorus~\cite{black}. Also, a bulk conductive layer of silicon (p- or n-type) can be used as a gate electrode, separated from the contact pattern by a layer of oxidized silicon~\cite{gete2w}.

\section{Superconducting proximity effect}

\subsection{Josephson current distribution}

Suppression of the critical current of an SNS junction by a magnetic field can be used to highlight the contribution of topological surface states, as it was firstly demonstrated for two-dimensional topological insulators~\cite{HgTeJ,InAsGaSbJ}. The idea of the method is as follows. With a uniform distribution of superconducting current across a wide SNS junction, the so-called Fraunhofer pattern of critical current suppression by a magnetic field is observed. Specifically, in a magnetic field, a phase shift appears between different one-dimensional channels  in the direction perpendicular to the current. Interference between these quasi-one-dimensional current-carrying regions creates a Fraunhofer curve with varying magnetic field (at field values significantly below the critical value). In contrast, if the superconducting current is carried primarily by edge states (for example, due to their significantly longer coherence length), then a magnetic field perpendicular to the sample controls the phase shift between two one-dimensional weak links. This means that one can expect the characteristic pattern of equidistant oscillations of the critical current with approximately constant amplitude in a magnetic field, which is characteristicfor a double-junction SQUID. It is precisely this difference that was demonstrated in topological and non-topological regimes for topological insulators~\cite{HgTeJ,InAsGaSbJ}.

\begin{figure}[t]
\center{\includegraphics[width=\columnwidth]{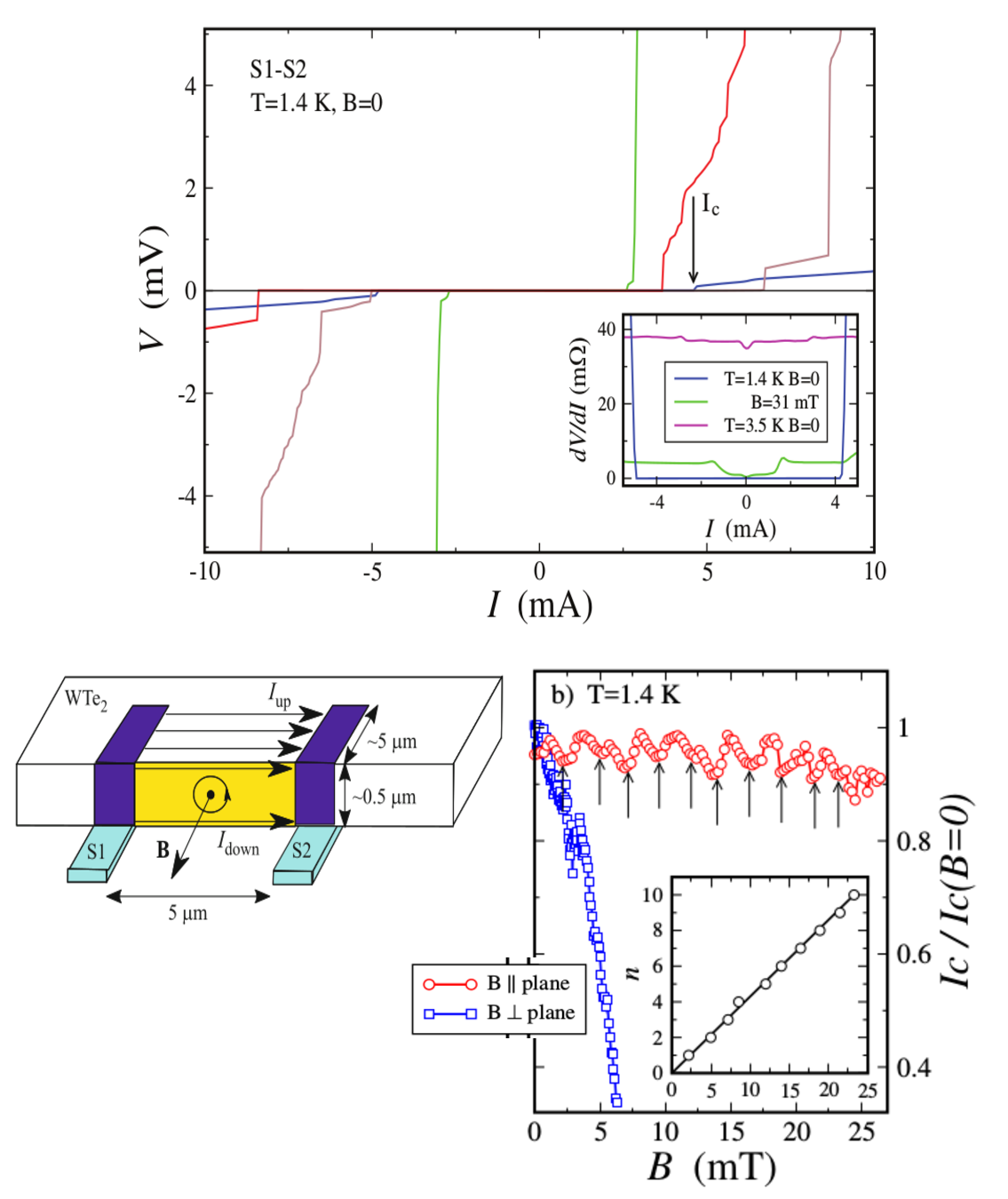}}
\caption{Top: Josephson current-voltage characteristics for several macroscopic (up to 5~$\mu$m) SNS junctions on the surface of the Weyl semimetal WTe$_2$. In the inset, the zero-resistance region is suppressed by temperature and magnetic field.
Bottom left: Effective sample geometry for the case of a parallel magnetic field (for a 5~$\mu$m junction, the bulk contribution can be neglected~\cite{NiTe2shapiro,hingeCdAs}). In a normal field, both surfaces are equivalent, which leads to a sharp drop in the critical current in a magnetic field, as expected for a single SNS junction. In a field parallel to the flake surface, a field-controlled phase shift occurs between the gaps, similar to a double-junction SQUID, which leads to equidistant oscillations of the critical current with approximately constant amplitude.
Bottom right: The pattern of critical current suppression for two  magnetic field directions, for the case of superconducting current transfer by topological states on two opposite surfaces of a thin WTe2 flake. The experimentally measured oscillation period (inset) in a parallel magnetic field corresponds to the area between the contacts. (From the papers~\cite{inWTe2in1,inWTe2in2})}
\label{WTe2squid}
\end{figure}

In the works~\cite{inWTe2in1, inWTe2in2, NiTe2shapiro}, Josephson current is demonstrated at macroscopic (down to 5~$\mu$m) distances along the surface of  the nonmagnetic Weyl semimetal~\cite{das16,feng2016} WTe$_2$, see Fig.~\ref{WTe2squid}  as an example. The magnitude of the critical current varies from sample to sample, but there is a clearly defined region of voltage zeroing at low currents, asymmetric (due to the direction of the current sweep), which reflects the dc Josephson effect. The region of zero resistance is suppressed by increasing the  temperature or the magnetic field. The very fact that the Josephson current flows over distances an order of magnitude greater than the estimated bulk coherence length in WTe$_2$ suggests superconducting current transfer by edge states, while the temperature dependence of the critical current in papers~\cite{inWTe2in1,inWTe2in2,NiTe2shapiro} corresponds to the ballistic transport regime~\cite{Tinkham}. Moreover, for Dirac semimetals, the contribution of hinge topological states to the transport properties was highlighted in works~\cite{NiTe2shapiro,hingeCdAs} precisely due to the sharply increased coherence length in such one-dimensional states compared to conventional surface states of the Dirac semimetal and bulk transport. Apparently, it is the topological protection of transport that is responsible for the increase in the coherence length~\cite{NiTe2shapiro,hingeCdAs}, which sharply distinguishes SNS junctions on the surface of topological semimetals from conventional metallic and semimetallic (bismuth, antimony) systems, where the Josephson current can only be observed at submicron lengths.

In work~\cite{inWTe2in1}, a 0.5~$\mu$m thick WTe$_2$ flake was heated to induce indium diffusion across the interface and establish a superconducting contact between two opposite flake surfaces. The pattern of the critical current suppression by the magnetic field begins to sharply depend on the magnetic field direction: the superconducting current is carried by topological states~\cite{das16,feng2016} on two sample surfaces (for a 5~$\mu$m junction, the bulk contribution can be neglected), i.e., we have two SNS junctions connected in parallel. In a normal field, there is no phase shift between these Josephson junctions, i.e., we observe a sharp drop in critical current in a magnetic field, as expected for each of the junctions, see Fig.~\ref{WTe2squid}. In a field parallel to the flake surface, a field-controlled phase shift occurs between the junctions, similar to a double-junction SQUID, which leads to equidistant oscillations of the critical current with approximately constant amplitude, see Fig. ~\ref{WTe2squid}. The effective geometry of the sample is shown in the inset to Fig.~\ref{WTe2squid}, the experimentally measured oscillation period corresponds to the area between the contacts~\cite{inWTe2in1}.

Importantly, this experimental setup eliminates the influence of possible artifacts, such as multiple indium shortings on the oxidized silicon surface. Even if we exclude the impossibility of indium diffusion over a distance of 5~$\mu$m, the presence of short circuits would lead to the appearance of a SQUID-like pattern of critical current suppression in a perpendicular, but not parallel, magnetic field. Thus, the experiment~\cite{inWTe2in1} is analogous to the earlier studies~\cite{HgTeJ,InAsGaSbJ} performed for two-dimensional topological insulators and demonstrates the possibility of superconducting current transfer over macroscopic distances by the surface states of a Weyl semimetal.

This result is confirmed by the observation of ac Josephson effect in paper~\cite{inWTe2in2}, where the presence of effective SQUID geometry is reflected in the appearance of fractional Shapiro steps~\cite{squid1,squid3}. The observation of ac Josephson effect in itself~\cite{inWTe2in2}, in addition to the vanishing of the dc resistance, completes the proof of Josephson current transport by surface states.

It should be noted that the implementation of a Josephson junction on the surface of a topological semimetal is conveniently achieved using indium as a superconducting contact material (see also the discussion below), but is not specific to this material. In the work ~\cite{NiTe2_flat}, niobium was used, and Josephson current was reliably demonstrated on the surface of the topological semimetal NiTe$_2$.

\subsection{Josephson current on the surface of magnetic topological semimetal}

Among the various implementations of Josephson structures based on topological semimetals, the possibility of superconducting current transport by surface states should be highlighted for the material with a ferromagnetic bulk. In particular, the Co$_3$Sn$_2$S$_2$ single crystal is a magnetic Weyl semimetal, the presence of topological surface states has been demonstrated experimentally using scanning tunneling spectroscopy~\cite{STM_cosns} and ARPES~\cite{ARPES_cosns}. Moreover, in terms of bulk properties, this material is characterized by full spin polarization (half-metal) in the single-domain state. Under these conditions, it is impossible to expect Josephson current flow through the bulk of the sample at micron distances. In contyrast, the transfer of Josephson current was demonstrated~\cite{cosns_SNS} along the surface of a thin flake of magnetic Weyl semimetal Co$_3$Sn$_2$S$_2$ .

\begin{figure}[t]
\center{\includegraphics[width=\columnwidth]{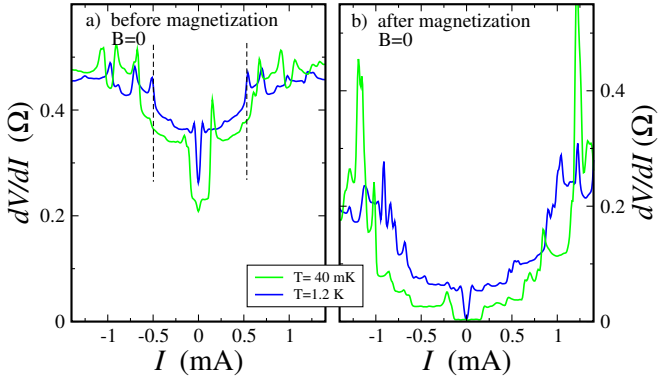}}
\caption{ Resistance between two superconducting contacts separated by 2~$\mu$m on the surface of the magnetic Weyl semimetal Co$_3$Sn$_2$S$_2$ at 1.2 K and 40 mK as a function of the current through the SNS junction. Left: magnetically disordered (multidomain) state immediately after cooling from above the Curie point in zero magnetic field. At voltages on the sample smaller than the superconducting gap, a decrease in the sample resistance is observed, as it would be expected for the Andreev reflection~\cite{andreev} for a relatively transparent contact~\cite{BTK}. Right: single-domain state of a Co$_3$Sn$_2$S$_2$ single crystal after magnetization to saturation; the topological surface state links two superconducting contacts separated by 2~$\mu$m, the bulk of the  crystal is completely spin-polarized. At low currents, a region of zero resistance appears, being well developed at 40 mK, which is a fingerprint of the Josephson current. Thus, an unambiguous connection between the presence of the anomalous Hall effect regime~\cite{cosns_AHE,cosns_AHE1} and the flow of Josephson current at micron distances is demonstrated, which is a clear demonstration of the transfer of Josephson current by spin-polarized topological surface state in the Co$_3$Sn$_2$S$_2$ single crystal (from the work~\cite{cosns_SNS})}
\label{CoSnS_JJ}
\end{figure}

After cooling the sample in a zero magnetic field, it is in a magnetically disordered state (the Curie point for this material is approximately 180 K). When measuring the resistance between two superconducting contacts separated by 2~$\mu$m, the resistance is finite for any current through the sample, and, accordingly, the voltage across the sample, see Fig. ~\ref{CoSnS_JJ}. At voltages lower than the superconducting gap, a decrease in the sample resistance is observed, as would be expected for the Andreev reflection~\cite{andreev} for a relatively transparent contact~\cite{BTK}, the effect depends weakly on temperature between 1.2 K and 30 mK. Thus, with a characteristic micron domain size in the disordered state of Co$_3$Sn$_2$S$_2$, the domain walls prevent direct connection between the contacts through the current-carrying topological state on the sample surface.

The resistance behavior changes dramatically after magnetizing the sample to saturation and removing the magnetic field. Co$_3$Sn$_2$S$_2$ is then in a single-domain state, as is known from observations of the anomalous Hall effect~\cite{cosns_AHE,cosns_AHE1} and direct magnetometric measurements~\cite{cosns_magnetization}. In this case, the topological surface state spreads over the entire surface of the sample (which is the origin of the anomalous Hall effect in Weyl semimetals~\cite{weyl_review}), in particular, it connects two superconducting contacts separated by 2~$\mu$m. The resistance of the sample begins to exhibit zero resistance at low currents, see Fig.~\ref{CoSnS_JJ}, which is especially clearly visible as a region of zero resistance of finite width at temperatures below 100 mK.

 Due to (i) the impossibility of superconducting current transport by the spin-polarized sample bulk, (ii) the demonstration of an unambiguous relationship between the presence of the anomalous Hall effect regime and the Josephson current flow at micron distances, the work~\cite{cosns_SNS} presents a clear demonstration of the Josephson current transport by a spin-polarized topological surface state, which is controlled by the magnetic state of the sample (multidomain -- single-domain). Such behavior is impossible for conventional metallic systems, which both highlights the role of surface states in transport in topological semimetals and excludes possible experimental artifacts (sample defects, etc.). In the work~\cite{cosns_SNS}, the suppression of the critical current by a magnetic field and temperature is studied, the dependences characteristic of the Josephson current  in a ballistic SNS junction~\cite{Tinkham} are shown.

\begin{figure}[t]
\center{\includegraphics[width=\columnwidth]{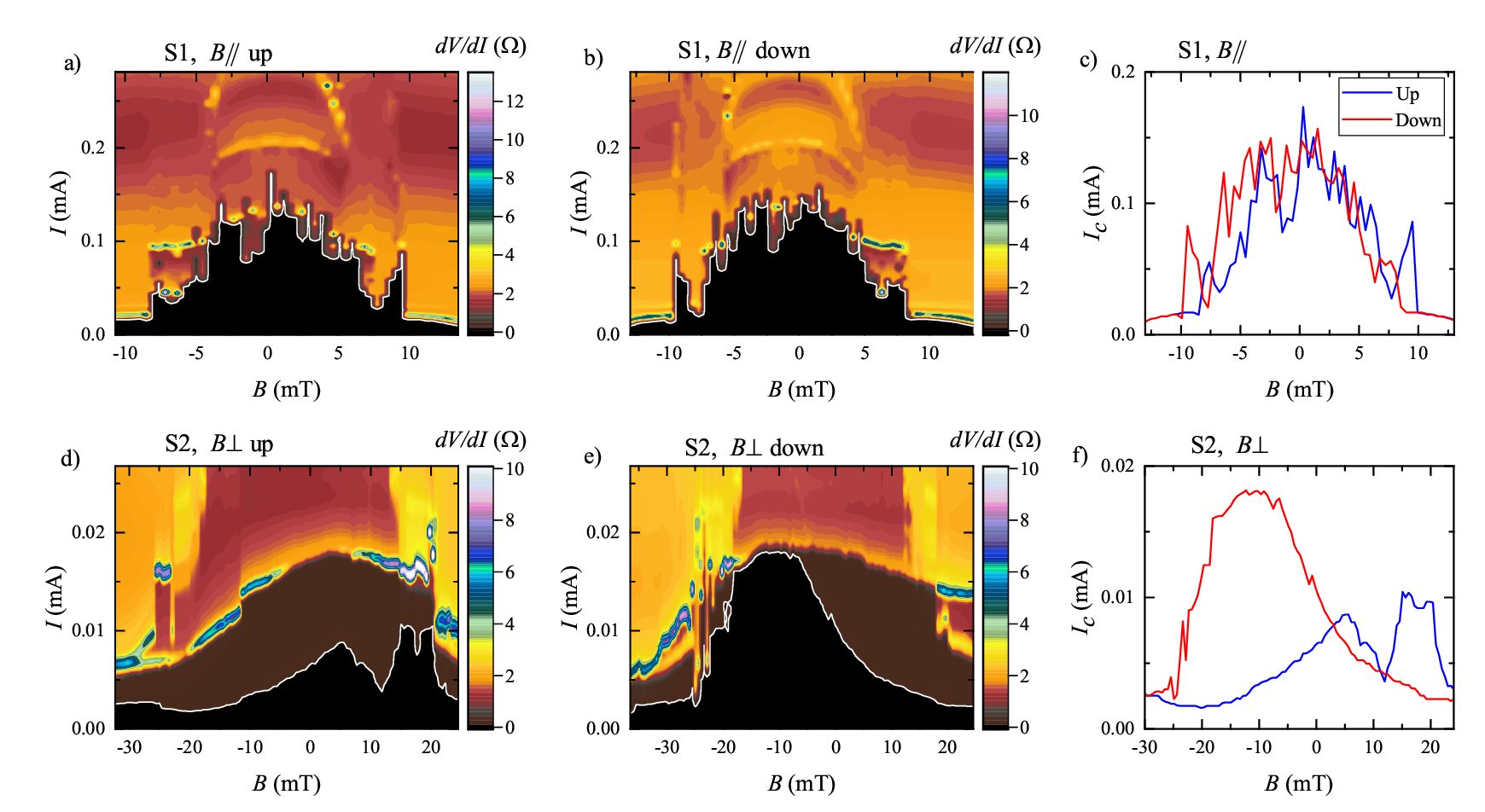}}
\caption{Asymmetry of the critical current of a Josephson junction on the surface of a magnetic nodal-line topological semimetal Fe$_3$Ge$_2$ , determined by the direction of the magnetic field sweep. The effect is most evident in a parallel field (lower panels), reflecting behavior characteristic of a  Josephson spin valve. Traces of this mirror asymmetry also persist for a perpendicular magnetic field orientation (from the paper~\cite{FGT_SNS}).}
\label{FGT_JJ}
\end{figure}

Already in the work~\cite{cosns_SNS}, the dependence of critical current suppression by a magnetic field was shown to be asymmetric with respect to zero field. This behavior was studied in detail~\cite{FGT_SNS} for the magnetic topological nodal-line semimetal~\cite{weyl_review} Fe$_3$GeTe$_2$ (FGT). It was shown that the detected asymmetry is determined by the direction of the magnetic field sweep, which is characteristic of a Josephson spin valve, see Fig.~\ref{FGT_JJ}. Typically, a spin valve is a ferromagnetic multilayer, where the electrical resistance is determined by the mutual orientation of the layer magnetizations due to spin-dependent scattering. In a Josephson spin valve, a ferromagnetic multilayer is sandwiched between two superconducting electrodes~\cite{krasnov}. In this case, the superconducting current is determined by the mutual orientation of the magnetizations, and not simply by the total magnetic flux, as for conventional Josephson junctions. The implementation of a spin valve in magnetic topological semimetals, associated with the presence of two spin-ordered subsystems, will be described below. In the work~\cite{FGT_SNS}, it was demonstrated that an SNS junction on the surface of a magnetic topological semimetal is a natural realization of a Josephson spin valve. 

\subsection{Single Andreev contact}

While studying the Josephson current flow along the surface of a topological semimetal provides information on charge transport by surface states, studying a single Andreev junction can reveal significant features of these states.

On the one hand, a single SN contact of a Weyl semimetal, as a well-conducting system, and a conventional (s-type) superconductor should demonstrate  the effect of Andreev reflection~\cite{andreev} on the SN interface. The latter can be seen as a decrease in resistance at voltages  within the superconducting gap for a relatively transparent interface~\cite{andreev} and its increase in the case of a highly disordered SN interface~\cite{BTK}. Thus, one should expect that the superconducting gap defines the only energy scale manifested in the current-voltage characteristics of the SN junction, and obtaining additional information specifically about the normal side of the junction is impossible. For topological semimetals, however, the presence of surface states at the SN interface leads to various subgap features in the current-voltage characteristics.

In the work~\cite{WTe2_SN} the transport was experimentally investigated through the interface between the Weyl semimetal WTe$_2$ and superconducting Nb. In the spectra of differential resistance $dV/dI(V)$ against the background of the standard Andreev signal, non-periodic resonances were detected inside the superconducting gap in niobium. From the analysis of their positions, their evolution with changing magnetic field and temperature, these resonances were interpreted as an analogue of Tomash geometric oscillations~\cite{Adroguer,Kopnin} due to the superconductivity induced in the topological surface state near the Nb-WTe$_2$ interface. Since the states in Fermi arcs  are supposed to have a specific group velocity direction~\cite{Okugawa} on the surface of a Weyl semimetal, the observation of distinct geometric resonances for an arbitrary contact shape confirms the presence of a preferred direction of charge motion in  transport experiment, which is apparently the only experimental confirmation of the chirality of surface states~\cite{WTe2_SN}.

Physics becomes more sophisticated for the proximity of a magnetic Weyl semimetal and a spin-singlet superconductor. Proximity-induced superconductivity in surface states is predicted even in the suppressed superconductivity regime in the bulk of a Weyl semimetal~\cite{Bovenzi}. For the case of triplet pairing, the appearance of a chiral Majorana mode is predicted at zero energy~\cite{Faraei}.

\begin{figure}[t]
\center{\includegraphics[width=\columnwidth]{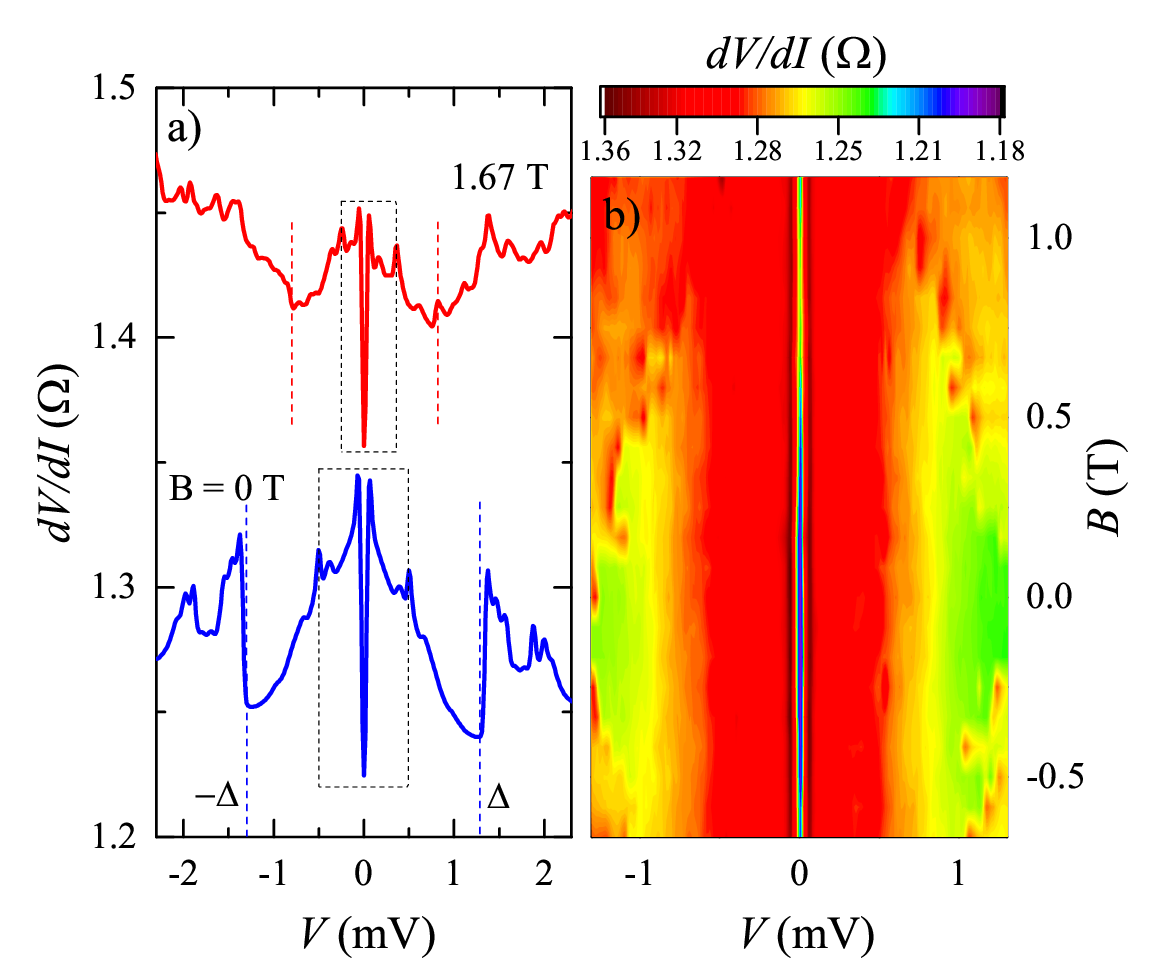}}
\caption{ Left: Differential resistance as a function of voltage at the Co$_3$Sn$_2$S$_2$-Nb interface. The zero-bias anomaly in the current-voltage characteristics of the interface is stable withing a significant range of the magnetic field. The superconducting gap (dashed line) is partially suppressed by the magnetic field, as expected for a superconducting niobium. Right: Demonstration of the stability of the zero-bias anomaly as a gradient plot of the differential resistance in the voltage-magnetic field axes (from the work~\cite{cosns_SN})}
\label{Nb_CoSnS}
\end{figure}

For a single superconducting contact, in work~\cite{cosns_SN} the charge transport was experimentally investigated through the interface between a superconductor (niobium) and a magnetic Weyl semimetal Co$_3$Sn$_2$S$_2$, which is characterized by complete spin polarization in the bulk. Along with the induced superconducting gap, a significant zero-bias anomaly was detected in the current-voltage characteristics. The observed zero-bias anomaly exhibits unusual stability with respect to an external magnetic field: the anomaly width remains constant up to the critical field in niobium, while the anomaly depth exhibits a weak nonmonotonic change in the magnetic field (see Fig.~\ref{Nb_CoSnS}).

This result was interpreted as   merging of Andreev states at zero energy at the interface, arising due to the proximity effect, under the influence of strong spin-orbit interaction and significant Zeeman splitting in spin-polarized (half-metallic) Co$_3$Sn$_2$S$_2$. This result is somewhat analogous to the merging of Andreev levels in one-dimensional nanowires in a magnetic field~\cite{Sarma}. In Weyl semimetals, the surface state is two-dimensional, but there are theoretical predictions extending the picture~\cite{Sarma} to the two-dimensional case~\cite{Woods}, facilitated by the effective one-dimensionality of chiral Weyl topological states.

A complete picture of the evolution of Andreev levels leading to their merging is demonstrated in the work ~\cite{CoSi_SN} for the topological halfmetal CoSi. This material is a realization of a chiral semimetal, where both the inversion symmetry (as in conventional Weyl semimetals) and mirror symmetry are broken~\cite{bernevig,zhang,chang2017}, and is characterized by a significant splitting of  Weyl nodes and, correspondingly, extremely long Fermi arcs~\cite{long,long1}. There have been demonstrated  moving of the Andreev levels to zero energy with increasing magnetic field, their merging at a certain critical field, and the stability of the zero energy level~\cite{CoSi_SN} at higher fields. However, based on the available data, it is impossible to assert the presence or absence of a topological transition leading to the emergence of a Majorana fermion~\cite{Sarma}.

\subsection{Reentrant superconductivity in SN and SNS structures based on the nonmagnetic topological semimetal GeTe}

Particular interest in non-magnetic GeTe is associated, among other things, with the giant (record-breaking) Rashba splitting known for this topological semimetal~\cite{ortix,triple-point} both for the bulk material~\cite{GeTerashba} and for surface states~\cite{GeTesurfStates}, which leads to the formation of complex spin textures~\cite{triple-point,spin text,GeTe_magnetization}. The results for GeTe deserve a separate section due to its suitability for practical applications (both the specific bulk spectrum and the surface states survive up to room temperature, see the section on the non-linear Hall effect below).

In the work~\cite{GeTe_In} the charge transport was experimentally investigated in NS and SNS proximity structures  based on non-magnetic topological semimetal GeTe. Nonmonotonic effects of an external magnetic field were observed for both types of structures, including reentrant superconductivity in the In-GeTe-In Josephson junction: as the magnetic field increases, the superconducting current disappears and reappears at higher magnetic fields.

\begin{figure}[t]
\center{\includegraphics[width=\columnwidth]{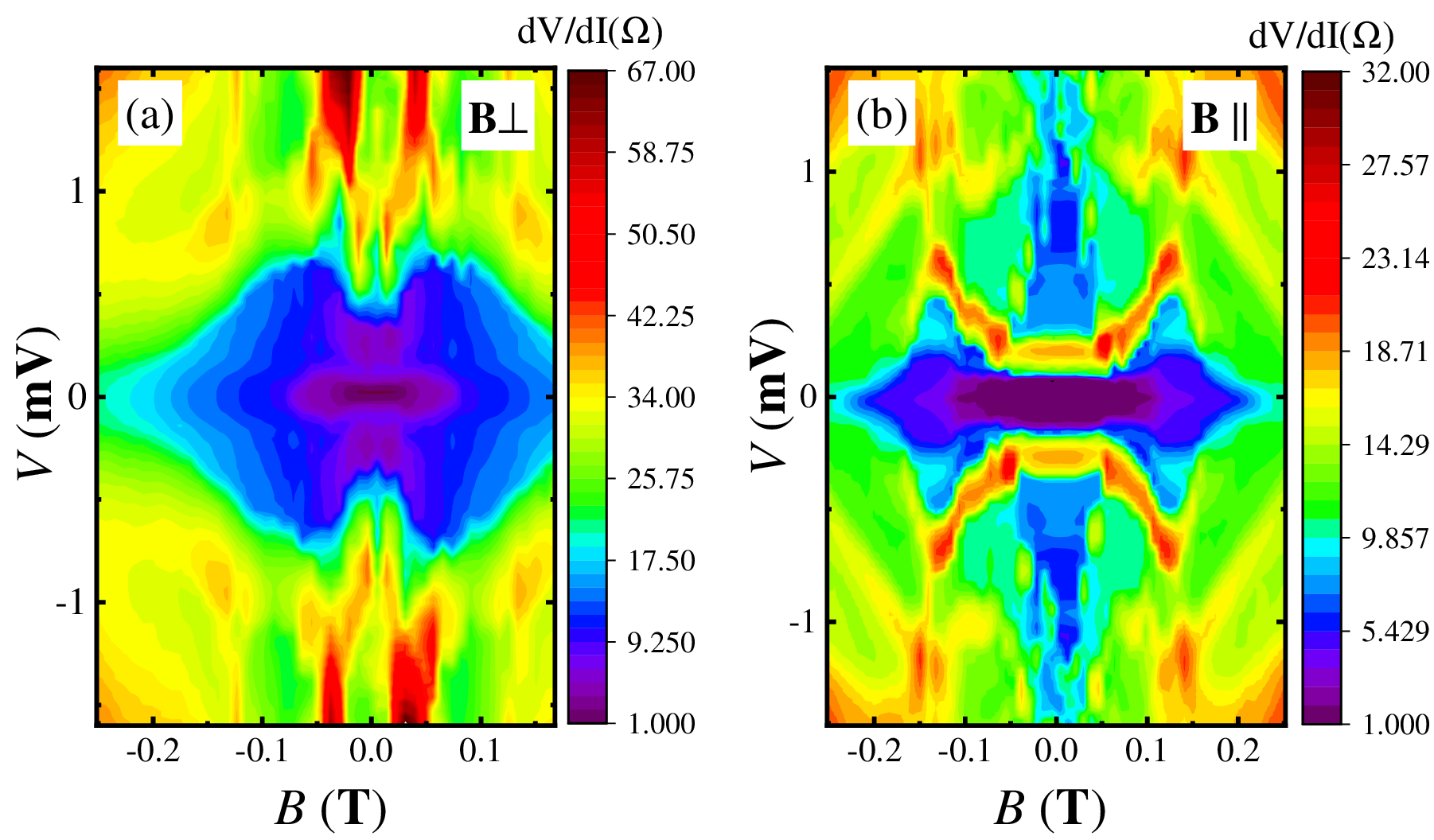}}
\caption{ Gradient plot of the differential resistance of a single In-GeTe junction shown in the voltage -- magnetic field axes. The superconducting gap is somewhat suppressed even in zero magnetic field, but it increases to a value characteristic of bulk indium in a certain range of magnetic fields before being completely suppressed by higher magnetic field. The effect is caused~\cite{BBK2002,Sperstadetal2008,Mengetal2013} by the non-uniform spin polarization formed by spin textures~\cite{spin text} in GeTe. Specifically, the magnetic field disrupts the initial spin configuration, causing the superconducting gap to increase to its bulk value, and then suppresses the gap in stronger magnetic fields. (from the work~\cite{GeTe_In})}
\label{GeTe_sns_sn}
\end{figure}

In the case of a single In-GeTe Andreev junction, the superconducting gap is already somewhat suppressed at zero magnetic field, but it increases to a value characteristic of indium in a certain range of magnetic fields before being completely suppressed by  higher magnetic field (see Fig. ~\ref{GeTe_sns_sn}). Since indium is a conventional s-type superconductor, the observed effects should be primarily related to the specific properties of GeTe. First of all, topological surface states with nontrivial spin textures may play an important role.~\cite{spin text} 

Thus, it seems quite natural to think of In-GeTe heterostructures as SFN junctions with a certain magnetization and strong spin-orbit coupling. The inhomogeneous spin polarization, which is believed to be incorporated into such spin textures~\cite{spin text}, complicates the model and leads to nonmonotonic behavior in magnetic fields~\cite{BBK2002,Sperstadetal2008,Mengetal2013}.
Specifically, a magnetic field disrupts the initial spin configuration, leading to an increase in the superconducting gap to its bulk value, and only then suppresses the gap in stronger magnetic fields. 

In the case of a planar Josephson In-GeTe-In geometry, supercurrent flows along the GeTe surface between superconducting indium contacts. In accordance with the behavior of the superconducting gap, the initial spin polarization of the surface states partially suppresses the critical current, which is restored due to modification of the surface spin structures by an external magnetic field, and is later completely suppressed in stronger magnetic fields. 

Thus, in the work~\cite{GeTe_In} reentrant superconductivity was associated with the inverse proximity effect in In-GeTe-In and In-GeTe structures.

\subsection{Concluding remarks on proximity-induced superconductivity}

It should be noted that all of the above-mentioned SNS structures based on magnetic and non-magnetic topological semimetals have demonstrated Josephson current transfer at macroscopic (up to 5~$\mu$m) distances along the surface of the topological semimetal.

The  Josephson current at distances an order of magnitude greater than the estimated bulk coherence lengths suggests superconducting current transport by topological surface states, with non-monotonic effects depending on the magnetic field emphasizing the presence of significant spin polarization of the surface states.

\section{Spin-dependent scattering}

As follows from both the initial theoretical concepts and from the experiments on spin-dependent ARPES, Fermi arcs in Weyl semimetals have significant spin polarization caused by spin-momentum locking~\cite{weyl_review,Xu,das16,feng2016}. This raises the question of whether such spin polarization can manifest itself in a direct transport experiment.

For ferromagnetic metal films, the spin valve  has been known since the early 2000s~\cite{myers,tsoi1,tsoi2}. It consists of two ferromagnetic layers of different thicknesses, separated by a normal metal layer. The current across such a multilayer is studied, and a lithographically created constriction~\cite{myers} or a point contact~\cite{tsoi1,tsoi2} is used to increase the current density. The physics of the phenomenon is as follows: when current flows through a thick layer, electrons are spin-polarized. When entering a thin one, the scattering processes, and therefore the sample's resistance, depend on the relative orientation of the layers magnetizations.

At low current densities, the resistance of a spin valve can be controlled by an external magnetic field: in a strong field, the layer magnetizations are necessarily parallel. As the field decreases to zero and its sign changes, layers of different thicknesses undergo magnetization reversals at different external field values, meaning an antiparallel configuration is possible over a certain field range. However, at higher fields, the layer magnetizations again become parallel. Thus, the resistance of the sample depends on the magnetic field, this dependence depends on the history (direction of the field sweep), that is, hysteresis is observed in the resistance of the sample.

Of particular interest is the case of oppositely oriented layer magnetizations at high current densities. In a zero magnetic field, spin-polarized electrons in the thick layer create a torque on the magnetization vector in the thin one, which can cause a reversal of the magnetization at a sufficiently high current density~\cite{myers}. In a magnetic field, even before the switching current is reached, the magnetization vector of the thin layer begins to precess around its equilibrium value, i.e., a spin wave is excited. In other words, spin-polarized electrons in the thick layer can not be regarded as proper excitations in the thin one, and  they pass through the thin layer without scattering. However, at a certain current density, magnon excitation is possible, which creates a new dissipation channel and manifests itself as an increase in the resistance of the multilayer. When measuring the differential resistance (which is more convenient in the experiment), the threshold excitation of the magnon appears as a sharp peak in the dependence of the differential resistance on the applied current~\cite{myers,katine}. Such peaks are asymmetric~\cite{myers,tsoi1,tsoi2} with respect to the polarity of the applied current due to the general asymmetry of the system (the presence of a thin and thick layers), which served as the basis for introducing the term spin valve. The position of the peak depends linearly on the magnetic field~\cite{myers,tsoi1,tsoi2}, which is well described~\cite{katine} by the Slonczewski model~\cite{slonczewski}. In a somewhat simplified version (without taking into account anisotropy), within the framework of this model, the critical current for excitation of the magnon~\cite{katine}
\begin{equation}
I_{sw}(B) \sim \alpha \gamma e \sigma B, 
\end{equation}
where $\alpha$ is the damping parameter, $\gamma$ is the gyromagnetic ratio, $e$ is the electron charge, and $\sigma$ is the total spin of the thin layer. In the single-domain regime, $\sigma$ is constant, which leads to a linear dependence of the position of the peak $I_{sw}$ on the magnetic field $B$.

\begin{figure}[t]
\center{\includegraphics[width=0.8\columnwidth]{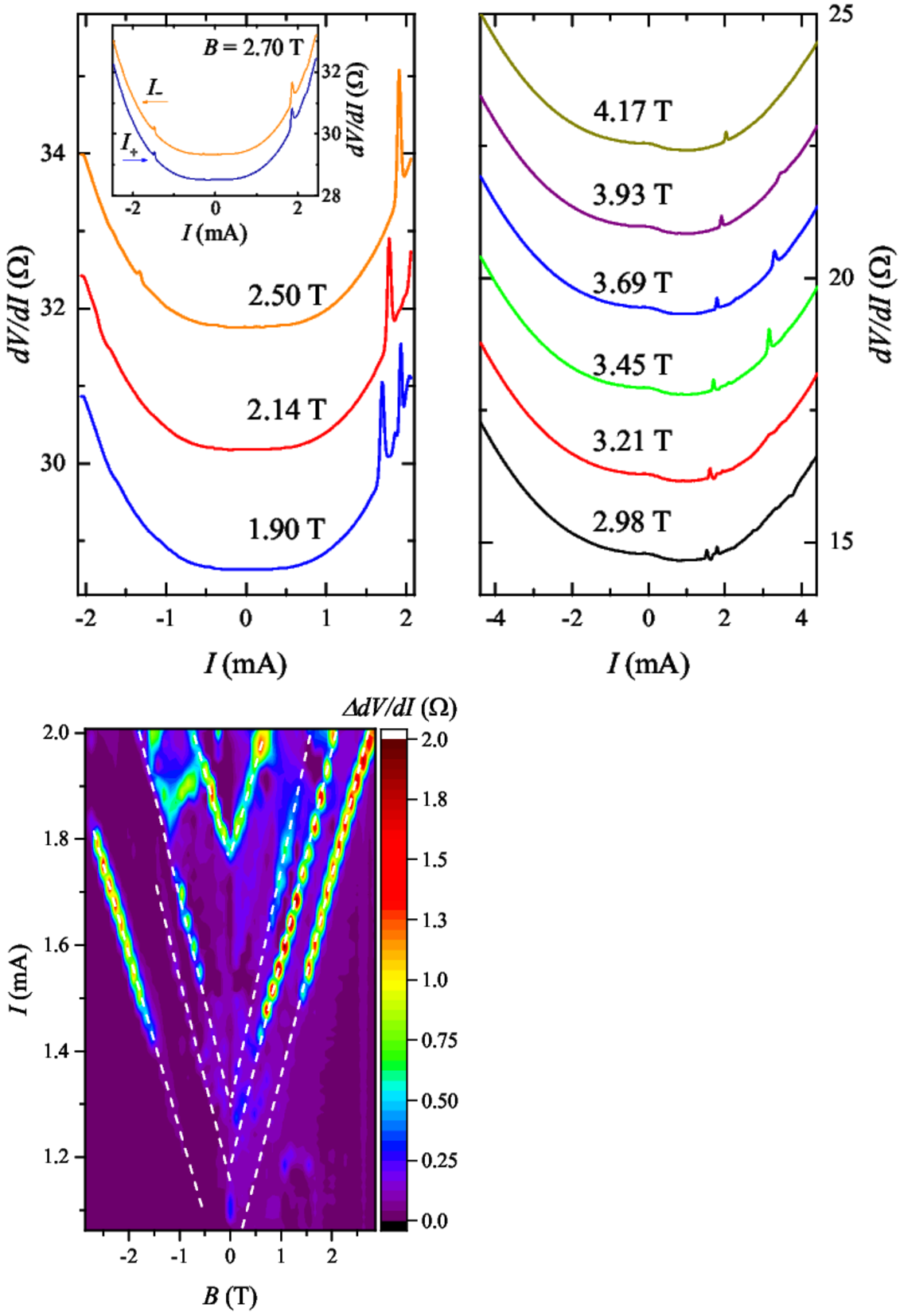}}
\caption{ A normal contact to a magnetic topological semimetal as a natural implementation of a spin valve. Two spin-polarized subsystems are realized as the ferromagnetic bulk of the crystal and spin-polarized (due to spin-momentum locking) topological surface states. Electrons acquire spin polarization during propagation in the topological surface state, and magnons are excited in the ferromagnetic bulk of the Co$_3$Sn$_2$S$_2$ single crystal.
Top row: differential resistance versus current through the contact, nonlinear characteristic shows asymmetric peaks in $dV/dI$ corresponding to magnon excitation. The inset demonstrates the absence of hysteresis when changing the current sweep direction.
Bottom left: linear dependence of the peak position in $dV/dI$ on the magnetic field strength, as required by the Slonczewski model~\cite{katine,slonczewski} for a spin valve. (From the work~\cite{cosns_magnon})}
\label{magnon}
\end{figure}

A magnetic topological semimetal is a natural realization of a spin valve~\cite{cosns_magnon,timnal_magnon,cosi_magnon,gete_spin_valve}. Unlike the artificial thick and thin layers in ferromagnetic multilayers, in a magnetic topological semimetal, two spin-polarized subsystems are realized as the ferromagnetic crystal bulk and spin-polarized (due to spin-momentum locking) topological surface states. When studying transport through a single (non-magnetic, e.g., gold) contact, the current spreading in the contact region will flow both through the surface states and within the bulk of the sample. Studying the transport properties of a single contact allows to realize high current densities.

In the work \cite{cosns_magnon}, the differential resistance of a single gold contact to the magnetic Weyl semimetal Co$_3$Sn$_2$S$_2$ was investigated at liquid helium temperatures (see Fig.~\ref{magnon}). The resulting current-voltage characteristic is highly nonlinear at high current densities (approximately $10^4$ A/cm$^2$). This nonlinearity manifests itself as an increase in the differential resistance at high currents, which rules out heating effects as a cause of the nonlinearity. In strong magnetic fields, peaks in the differential resistance appear on the current-voltage characteristic for one current polarity, i.e., in exact agreement with the picture observed in ferromagnetic multilayers~\cite{myers}. The curves show no hysteresis with the direction of the current sweep, while the position of the peak in the differential resistance depends linearly on the magnetic field, as required~\cite{katine} by the Slonczewski model~\cite{slonczewski}. Unlike ferromagnetic multilayers, the magnetic Weyl semimetal Co$_3$Sn$_2$S$_2$ exhibits a set of magnon branches, with a complex interplay between them as the magnetic field changes (see Fig.~\ref{magnon}).

A reasonable interpretation is that electrons acquire spin polarization during propagation in the topological surface state (almost complete polarization, see as an example~\cite{Xu,das16,feng2016}), and magnons are excited in the ferromagnetic bulk of the Co$_3$Sn$_2$S$_2$ single crystal. Due to the complex structure of the topological surface states~\cite{STM_cosns,ARPES_cosns} in Co$_3$Sn$_2$S$_2$, multiple magnon modes are excited, while the gapless nature of the Weyl excitations in the topological semimetal is responsible~\cite{weyl_magnon} for the suppression of the damping~\cite{dirac_magnon} and, accordingly, for a sharp decrease in the current density required to excite the magnon~\cite{katine,slonczewski} (several orders of magnitude lower than in multilayers~\cite{myers,tsoi1,tsoi2}).

Similar experiments have been conducted for other topological semimetals. In particular, in the work~\cite{timnal_magnon} a direct experimental comparison of two systems was carried out: a normal (gold) contact to the magnetic Weyl semimetal Ti$_2$MnAl and a ferromagnetic (nickel) contact to the non-magnetic Weyl semimetal WTe$_2$. In this case, the systems under consideration have virtually nothing in common other than the presence of a topological spin-polarized surface state at the interface and a three-dimensional ferromagnet adjacent to the interface, realized either as a metallic nickel contact or as a ferromagnetic bulk of Ti$_2$MnAl. The choice of the not very popular Weyl semimetal Ti$_2$MnAl was due to its magnetic properties: in weak magnetic fields, Ti$_2$MnAl exists in a multidomain state, like nickel (a soft magnetic ferromagnet), making a direct comparison justified. For both studied systems, nonlinear behavior of the differential resistance was observed at high current densities through the contact, with the peaks of the differential resistance depending on the magnetic field in the same way. The behavior of the peaks in a magnetic field is more sophisticated than in single-domain Co$_3$Sn$_2$S$_2$ and can be explained by the motion of domain walls in the contact region~\cite{cosi_magnon}. However, what is most important is that the similarity of behavior of two completely different systems indicates the identical physics of the processes, which is expressed in the acquisition of spin polarization by electrons in a topological surface state and the excitation of magnons in a three-dimensional ferromagnet by a spin-polarized current~\cite{timnal_magnon}.  

Similar result can be shown for the chiral semimetal CoSi, characterized by the presence of surface ferromagnetism~\cite{cosi_magnon}. In addition to the symmetry-induced spin-orbit interaction~\cite{Burkovetal2018}, CoSi exhibits surface ferromagnetism of unknown origin, despite a nominally diamagnetic bulk. This allows the implementation of a spin valve in yet another system -- a normal (gold) contact to a bulk-nonmagnetic CoSi single crystal, where spin polarization of the current arises in the surface states, and the magnon is excited in the near-surface ferromagnetic region. The results are in perfect agreement with those obtained in~\cite{timnal_magnon}, which confirms the universality of the effect - all these systems have in common only the presence of the surface states.

Several remarks are worth making on the origin of surface ferromagnetism in CoSi. It has been theoretically predicted that a superlattice of ordered defects is a more probable source of CoSi ferromagnetism than dangling bonds at the surface~\cite{CoSiNW1,CoSiNW2,seo,tai}. In the work~\cite{cosi_magnetization}, the magnetization hysteresis for initially oxidized CoSi crystals and cleaved samples with a clean, oxide-free surface was compared experimentally. While oxidized CoSi samples did not exhibit a significant ferromagnetic response, the fresh CoSi surface demonstrated a strong ferromagnetic signal, accompanied by a pronounced modulation in the angular dependence of magnetization. This behavior emphasizes the role of the surface, it weakly correlates with the origin of ferromagnetism from ordered defects (vacancies) in the bulk of the material, and it is more consistent with the interaction of dangling bonds at the surface of the CoSi crystal~\cite{CoSiNW1,CoSiNW2,seo,tai}. However, according to theoretical calculations~\cite{seo,tai}, the latter is insufficient for the appearance of a ferromagnetic response, which requires taking into account the RKKY-type enhancement of the interaction by topological surface states of the chiral semimetal~\cite{Duan45,Zyuzin46,Duan2022}.
 
Thus, the works~\cite{cosns_magnon,timnal_magnon,cosi_magnon} demonstrated the universality of the assertion that a magnetic topological semimetal is a natural realization of a spin valve.

This phenomenon can also be exploited for practical applications. In the work~\cite{gete_spin_valve}, the spin-valve effect with a reversal of the magnetic field direction was demonstrated for a ferromagnetic (nickel) contact to the topological semimetal GeTe at low current densities. Hysteresis in the resistance of such a contact was demonstrated, including at room temperature, in contrast to the systems discussed above and, for example, the Ni-NiTe$_2$ system, where liquid helium temperatures and below are required to demonstrate the spin-valve effect~\cite{gete_spin_valve}. In this regard, the GeTe topological semimetal appears to be the most suitable for practical applications, see also the results for the nonlinear Hall effect below.

\section{Magnetic response of topological surface states}

Since  (i) the observation of the spin-valve effect~\cite{cosns_magnon,timnal_magnon,cosi_magnon,gete_spin_valve} for a single crystal of a magnetic topological semimetal demonstrates the presence of a second spin-polarized subsystem in the material in addition to the spin-polarized bulk, and (ii) the spin polarization of the topological surface state is confirmed by ARPES~\cite{weyl_review,Xu,das16,feng2016}, the question arises about the possibility of detecting the magnetic response of surface states in direct magnetometric measurements.

\begin{figure}[t]
\center{\includegraphics[width=\columnwidth]{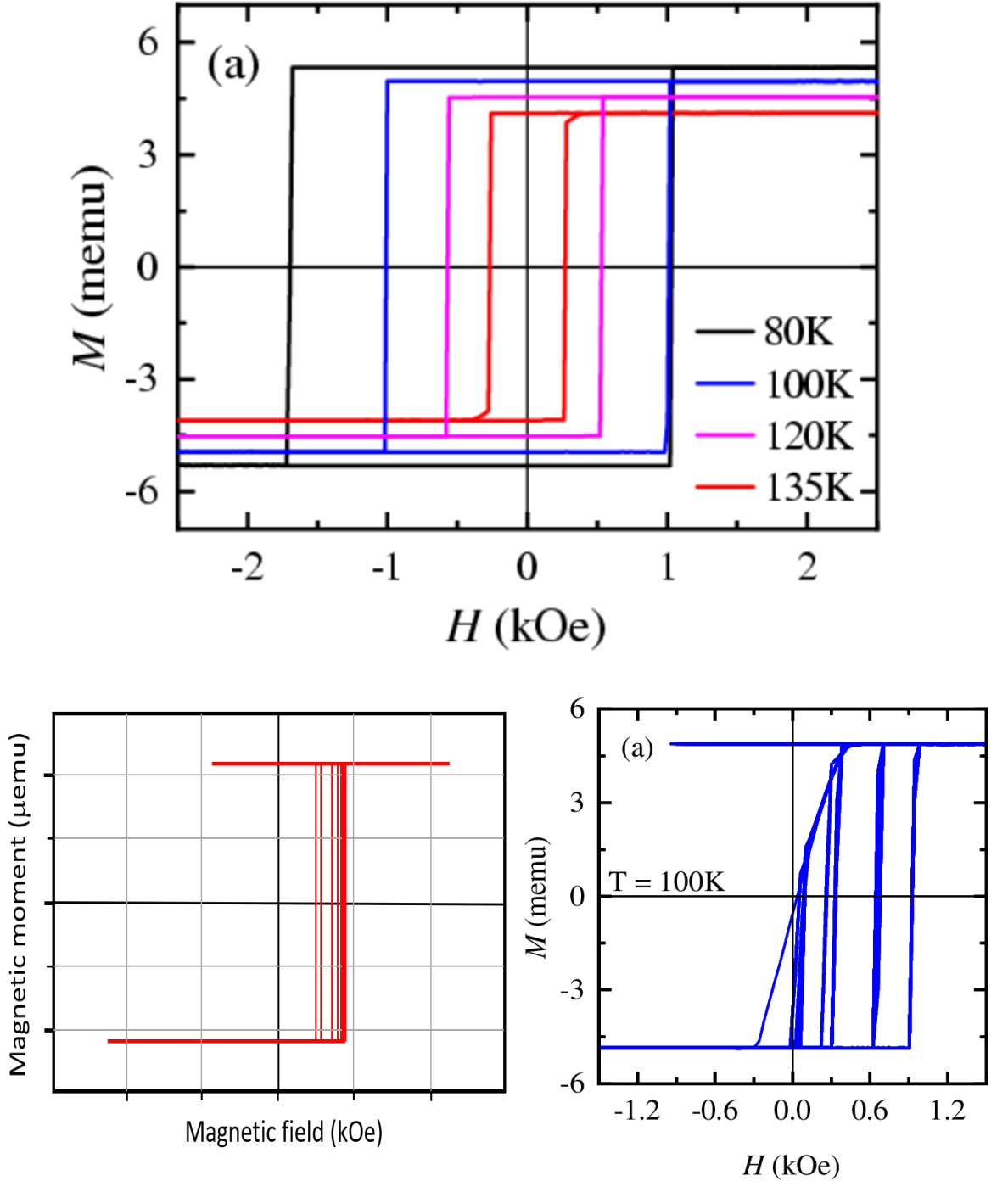}}
\caption{Top: magnetization hysteresis in the form of a rectangular loop at temperatures below 135 K for the magnetic Weyl semimetal Co$_3$Sn$_2$S$_2$, confirming a single-domain ferromagnetic state. Bottom left: expected behavior of the first-order magnetization reversal curves $M(H_r,H)$ for a rectangular hysteresis loop. Bottom right: experimental FORC curves, supporting the multi-phase behavior  even at 100 K: in addition to multiple step-like switchings at positive magnetic fields, there is an additional set of FORC curves with the finite slope,  corresponding to the presence of a second magnetic phase.  Moreover, FORC measurements demonstrated high temperature stability of this phase, in contrast to conventional bulk ferromagnetism in these materials. This temperature stability, as well as the excellent reproducibility of this result for two different topological semimetals Co$_3$Sn$_2$S$_2$ and Fe$_3$GeTe$_2$, allows us to relate the second phase to the magnetic response of spin-polarized topological surface states. (From the work~\cite{cosns_magnetization})}
\label{magnetization}
\end{figure}

In the work~\cite{cosns_magnetization} the presence of a contribution of surface states in the magnetic response was demonstrated at temperatures well below the Curie temperature in the topological semimetals Co$_3$Sn$_2$S$_2$ and Fe$_3$GeTe$_2$. Both conventional magnetization reversal curves (magnetization hysteresis when changing the direction of the magnetic field sweep between two saturation fields of opposite sign) and the comparatively rarely used technique of studying first-order magnetization reversal curves (FORC) were investigated. For conventional magnetization reversal curves, the presence of a rectangular hysteresis loop was shown at temperatures below 135~K, i.e., the sample is in the single-domain state, see Fig.~\ref{magnetization}. At higher temperatures, the appearance of an inverted hysteresis loop has been demonstrated, which is usually considered as evidence of the coexistence of several magnetic phases (the presence of a bias field in one phase causes a switch in magnetization in the second even before the sign of the external magnetic field changes).

The mere presence of two phases above a certain temperature cannot be proof of a surface state response, since the surface states  exist at any temperature as long as the bulk spectrum is preserved. However, the use of first-order magnetization reversal curves (FORC) has shown that this second sufface-state-induced phase exists even at the lowest temperatures, see Fig. ~\ref{magnetization}.

Specifically, in FORC measurement, the sequence of steps can be described as follows: 1. The base value of the magnetic field $H_s$ is established, exceeding the saturation field $H_0$ of the given material. Since the saturation of magnetization is achieved, the exact value of the base field $H_s>H_0$ does not affect the result. 2. The magnetic field changes abruptly to a certain value $H_r<H_s$, starting from this value, the magnetization curve $M(H)$ is measured, with the field $H$ changing  from $H_r$ to $H_s$. Steps 1 and 2 are repeated for different values of $H_r$, which cover the interval from $H_s$ to $-H_s$ with a given increment $\Delta H_r$. For the obtained $M(H_r,H)$ dependencies, further processing is possible (taking the second derivative to obtain a gradient image, as described, for example, in the work~\cite{cosns_magnetization}), however, the $M(H_r,H)$ curves themselves are already informative enough to describe the details of the  magnetization process inside the hysteresis loop.  

In particular, for a rectangular hysteresis loop for Co$_3$Sn$_2$S$_2$ at low temperatures (below 135 K), one can expect a simple form of FORC curves, as shown in Fig.~\ref{magnetization}: $M(H_r,H)$ is a horizontal line for all values of $H_r>-H_0$, with the magnetization value corresponding to the positive saturation level. With a further decrease in $H_r$, the $M(H)$ curve describes the lower half of the hysteresis loop (possibly with a switching field $+H_0$ that does not immediately establish itself for values of $H_r\le -H_0$). This is precisely the behavior one should expect for a single-phase single-domain system.

In experiment~\cite{cosns_magnetization}, in addition to the described pattern, an additional slanted region appears on the $M(H_r,H)$ diagram, which might be expected for a soft magnetic (multidomain) sample, see Fig.~\ref{magnetization}. Thus, even at the lowest temperatures, two magnetic phases exist in the sample, but the second, corresponding to the slanted region on the FORC diagram, cannot be determined by conventional hysteresis loop measurements, since the width of this region is smaller than of the hysteresis loop one. Moreover, FORC measurements demonstrated the high temperature stability of this phase, in contrast to conventional bulk ferromagnetism in these materials~\cite{cosns_magnetization}. With increasing temperature, the second magnetic phase manifests itself as inverted hysteresis due to the narrowing of the rectangular loop corresponding to the single-domain bulk of the sample. This temperature stability, as well as the excellent reproducibility of this result for two different topological semimetals Co$_3$Sn$_2$S$_2$ and Fe$_3$GeTe$_2$, allows us to relate the second phase to the magnetic response of spin-polarized topological surface states.

\section{Nonlinear anomalous Hall effect (NLHE)}

Already in the introduction (\ref{berry_section}), we discussed that the effective magnetic field (Berry curvature) $F_n=\nabla_k \times A_n$ leads to the appearance of a Hall term $\sim eE\times F$ for the electron motion in real space. However, the Hall voltage does not arise in a system without breaking time-reversal symmetry: as shown in Fig.~\ref{Fermi-arcs-ARPES}, in a non-magnetic Weyl semimetal, the Fermi arcs point towards each other, so in zero magnetic field there is no net charge transfer by the edge states and, accordingly, no net Hall term. However, the situation changes if the electric field is directed along the $a$ axis of the crystal in Fig.~\ref{Fermi-arcs-ARPES}. Under these conditions, an imbalance arises between electrons flowing in opposite directions, which leads to a Hall signal in the second order in the electric field, since the difference in the effective magnetic fields becomes proportional to the electric field of the flowing current (Berry curvature dipole)~\cite{sodemann}.

\begin{figure}[t]
\center{\includegraphics[width=\columnwidth]{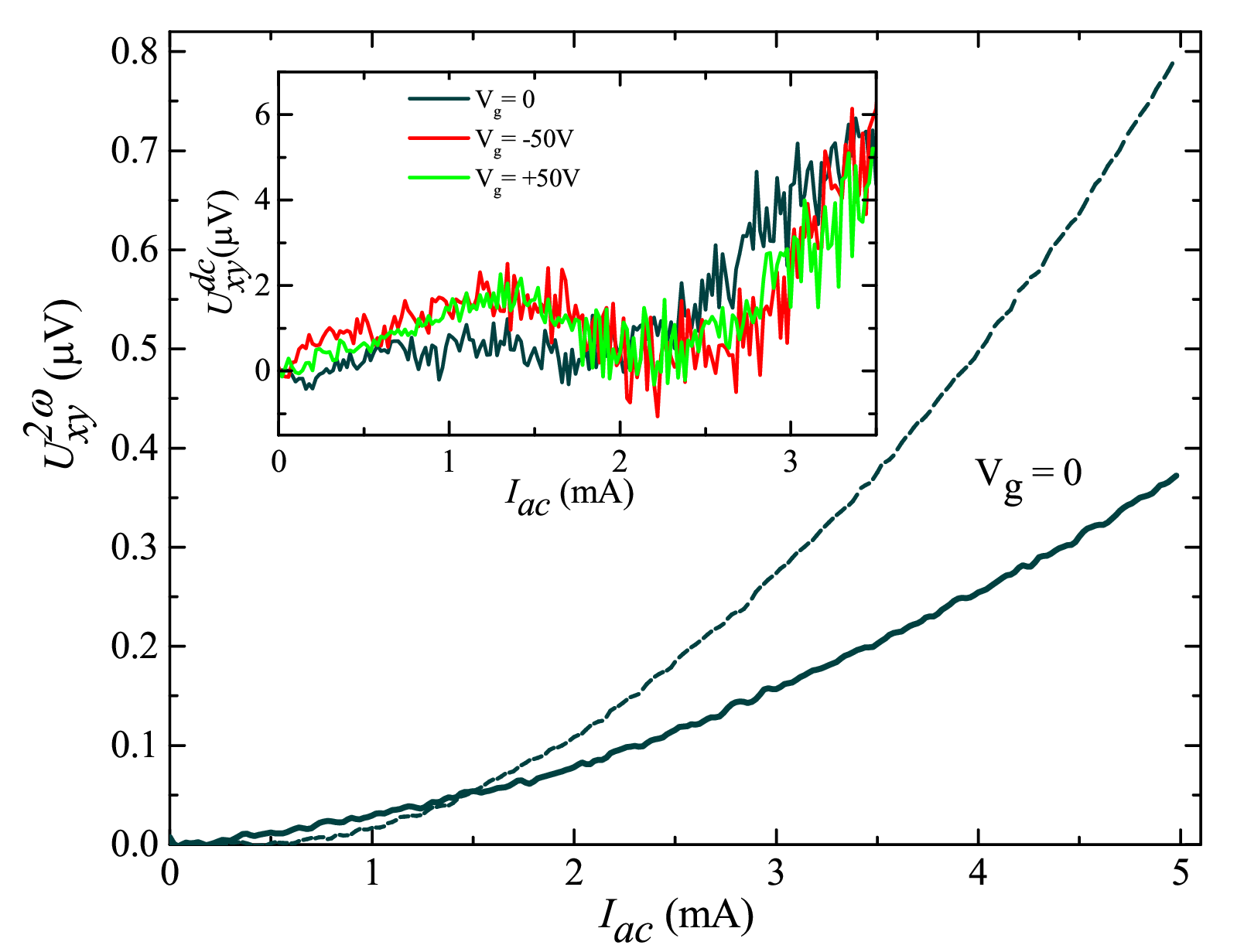}}
\caption{ Nonlinear anomalous Hall effect~\cite{sodemann} (NLHE) as a nonzero transverse (Hall) signal $V_H\sim I_0^2$ measured at the second $2\omega$ harmonic with respect to the  alternating longitudinal current $I_0\sin(\omega t)$ for two different samples. The inset shows the measurement result at the zero harmonic (DC) for comparison. One can see the distortion of the signal shape compared to a purely square-law signal at the second harmonic, a high noise level, but the very presence of a rectified signal controlled by the gate voltage has practical application perspective~\cite{tera}. The data were obtained for the topological semimetal GeTe at room temperature (from the work~\cite{gete2w}).}
\label{weyl2w}
\end{figure}

This effect is called the nonlinear anomalous Hall effect (NLHE). It has been predicted theoretically~\cite{sodemann} for several classes of systems without time-reversal symmetry breaking but with nonzero Berry curvature: crystalline topological semimetals, three-dimensional Weyl semimetals, and two-dimensional transition metal dichalcogenides.

The simplest way to study the nonlinear anomalous Hall effect is as a nonzero transverse (Hall) signal measured at the second $2\omega$ harmonic with respect to the harmonic alternating longitudinal current $I_0\sin(\omega t)$, see the example in Fig.~\ref{weyl2w}. In principle, due to the quadratic nature of the effect $V_H\sim I_0^2 \sin^2(\omega t) \sim I_0^2 (1+\cos(2\omega t))$, the NLHE signal, proportional to the square of the longitudinal current amplitude, can be measured either at the zero or the second harmonic. However, measurements at the second harmonic allow the use of lock-in  technique, which significantly improves the signal-to-noise ratio. It also eliminates various interferences (contact potentials, etc.). The inset to Fig. ~\ref{weyl2w} shows, for comparison, the measurement result at the zero harmonic (DC) when passing AC current through the sample. The distortion of the signal shape is evident, in comparison with a purely square-law signal at the second harmonic, as well as the high noise level. However, a non-zero dc Hall signal is reliably demonstrated for GeTe, even at room temperature~\cite{gete2w}. 

The nonlinear anomalous Hall effect was first demonstrated experimentally for two-dimensional transition metal dichalcogenides, and the correct spatial symmetry of the effect was demonstrated~\cite{ma,kang}. For three-dimensional systems such as Weyl semimetals, the effect was first demonstrated in the work~\cite{esin}, measurements in an external magnetic field made it possible to prove the nature of the effect (Berry curvature dipole). In particular, a serious experimental challenge is separating the contributions from the nonlinear anomalous Hall effect and the thermoelectric power.

When a current flows through the sample, Joule heating of the crystal occurs. Due to finite thermal conductivity, in an ideal geometry, the crystal temperature decreases symmetrically in the directions perpendicular to the current line. Thus, one can expect that in an ideal Hall geometry, both potential Hall probes are at the same temperature, and the thermal emf does not contribute to the measured signal. On the other hand, the real geometry can be non-ideal, despite the experimenter's best efforts. Under these conditions, a thermopower  signal (the Seebeck effect) appears, proportional to the temperature difference and, accordingly, to the Joule heating $V_H\sim R_{xx} I_0^2 \sin^2(\omega t)$, which naturally leads to the appearance of a signal at the second harmonic. It is impossible to distinguish a priori  these two contributions in  zero magnetic field, but they behave differently when a magnetic field is applied. The thermo-EMF signal is an even function of the field, since the magnetoresistance $R_{xx}(B)$ is insensitive to the field direction (the Nernst effect gives no income in the Hall geometry), while the signal of the nonlinear anomalous Hall effect is linear, and therefore odd, in the magnetic field~\cite{NLHE_linear_B,NLHE_linear_B1}.

\begin{figure}[t]
\center{\includegraphics[width=\columnwidth]{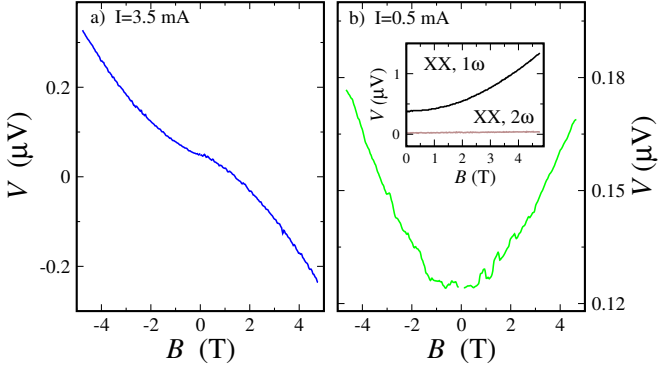}}
\caption{ Left: Odd dependence of the second-harmonic transverse voltage on magnetic field, becoming linear in strong fields, for the most ideal, symmetric, Hall geometry, as expected for the nonlinear anomalous Hall effect. Right: For contacts exactly perpendicular to the current line, but located at unequal distances from it, the measured even dependence of the second-harmonic transverse voltage on magnetic field exactly matches the magnetoresistance observed in the samples (shown in the inset as a longitudinal $V_{xx}$ signal at the first harmonic) and allows us to identify the signal as originating from the thermal emf. The results were obtained for the Weyl semimetal WTe$_2$ at a liquid helium temperature of 4.2 K (from the work~\cite{esin})}
\label{weyl2w_field}
\end{figure}

This qualitative difference in the dependence of the Hall voltage on the magnetic field was demonstrated in the work~\cite{esin} for two experimental geometries, see Fig.~\ref{weyl2w_field}. In the maximally ideal Hall geometry, an odd dependence of the second-harmonic transverse voltage on the magnetic field was shown, becoming linear in strong fields, which allows us to identify the Hall voltage as originating from the nonlinear anomalous Hall effect (it should be noted that the signal is nonzero already at zero magnetic field, in contrast to the usual Hall signal). For contacts exactly perpendicular to the current line, but located at an unequal distance from it, the dependence of the second-harmonic transverse voltage on the magnetic field is even, it exactly corresponds to the magnetoresistance observed in the samples (shown in the inset as a longitudinal $V_{xx}$ signal at the first harmonic). Moreover, all these experiments verify the absence of both the Hall signal at the first harmonic and the longitudinal signal at the second harmonic in zero magnetic field. Thus, in the work~\cite{esin} a method for reliably identifying the signal of the nonlinear anomalous Hall effect is demonstrated.      

Similar experiments have been performed for various materials~\cite{c_axis,cosi2w,fgt2w,gete2w}. GeTe deserves special attention, since the nonlinear anomalous Hall effect is observed already at room temperature~\cite{gete2w}, see Fig.~\ref{weyl2w}, due to the peculiarities of the band structure of this material, in particular, the giant Rashba splitting~\cite{GeTerashba} known for this topological semimetal~\cite{ortix,triple-point}. The latter, among other things, can be controlled by an external electric field at a constant carrier concentration~\cite{GeTerashba,spin text,GeTeour}, which makes it possible to realize a gate-field-controlled nonlinear anomalous Hall effect~\cite{gete2w}. The influence of the gate field was also demonstrated for two-dimensional dichalcogenides WTe$_2$~\cite{NLHEgate}, although the reason for the effect in this case is more trivial - in the two-dimensional case there is no problem with screening of the gate field by carriers in the sample, unlike three-dimensional GeTe~\cite{GeTeour}.

On the one hand, controlling the nonlinear anomalous Hall effect with an external electric field is useful for practical applications. In the case of a significant response at zero frequency~\cite{gete2w,NLHErect,NLHErect1}, the effect can be used to detect high-frequency signals, up to terahertz~\cite{tera}, without thermal losses. On the other hand, the dependence on the gate electric field emphasizes the role of the nonzero Berry curvature in the origin of the effect: it is difficult to expect the effect from the gate voltage, experimentally demonstrated in~\cite{gete2w,NLHEgate} for the contribution from asymmetric scattering by nonmagnetic impurities (skew scattering)~\cite{skew}.

Thus, the nonlinear anomalous Hall effect is a direct demonstration of non-zero Berry curvature in topological materials, obtained in a transport experiment.

\section{Conclusion. Possible applications.}

This review presents the experimental results on charge and spin transport in topological semimetals: charge transport in   different superconducting  proximity devices; spin-dependent transport; magnetic response of the topological surface states; non-linear anomalous Hall effect as the direct manifestation of the non-zero Berry curvature  in topological semimetals.  

From the point of view of practical applications, the most obvious ones can be assumed to be the use of the nonlinear anomalous Hall effect for the efficient detection of high-frequency signals~\cite{tera}, the use of the anomalous Hall effect for the magnetic memory~\cite{magnetoresisitve_memory}, and, as was shown experimentally for the magnetic chiral topological system~\cite{resistance_standard}, realization of a resistance standard that does not require  the use of an external magnetic field in contrast to the quantization of the Hall resistance in two-dimensional systems.

\section{Acknowledgments}
The author expresses gratitude to his colleagues: A.V. Timonina and N.N. Kolesnikov, without whose samples the experiments presented here would not have been possible; A.A. Kononov, O.O. Shvetsov, V.D. Esin, and N.N. Orlova, for long-term collaboration in experimental research; and Yu.S. Barash for theoretical support and wise advices.

\end{document}